\documentclass[journal]{IEEEtran}
\usepackage[english]{babel}
\usepackage{amsmath}
\usepackage{graphicx}
\usepackage[colorlinks=true, allcolors=blue]{hyperref}
\usepackage{xcolor}
\usepackage{comment}
\usepackage{lipsum}
\usepackage{lettrine}
\usepackage{cite}  
\usepackage{multicol}
\usepackage{makecell}
\usepackage{colortbl}
\usepackage{xcolor}
\usepackage{booktabs}
\usepackage{balance}

\title{Odor-Based Molecular Communications: State-of-the-Art,  Vision, Challenges, \\ and Frontier Directions}

\author{Dilara Aktas,~\IEEEmembership{Student Member,~IEEE}, Beyza E. Ortlek,~\IEEEmembership{Student Member,~IEEE}, Meltem Civas,~\IEEEmembership{Student Member,~IEEE}, Elham Baradari, Ayse S. Okcu, Melanie Whitfield, \\ Oktay Cetinkaya,~\IEEEmembership{Senior Member,~IEEE}, and~Ozgur~B.~Akan,~\IEEEmembership{Fellow,~IEEE}
\thanks{The authors are with the Center for neXt-generation Communications (CXC), Department of Electrical and Electronics Engineering, Ko\c{c} University, Istanbul 34450, Turkey (e-mail: dilaraaktas20@ku.edu.tr).}
\thanks{O. B. Akan is also with the Internet of Everything (IoE) Group, Electrical Engineering Division, Department of Engineering, University of Cambridge, Cambridge CB3 0FA, UK (email: oba21@cam.ac.uk).}
\thanks{This work was supported in part by the AXA Research Fund (AXA Chair for Internet of Everything at Ko\c{c} University) and Huawei Graduate Research Scholarship.}
}

\begin{document}
\bstctlcite{IEEEexample:BSTcontrol}
\setlength{\parskip}{0pt}

\maketitle

\begin{abstract}
Humankind mimics the processes and strategies that nature has perfected and uses them as a model to address its problems. That has recently found a new direction, i.e., a novel communication technology called molecular communication (MC), which utilizes molecules to encode, transmit, and receive information. Despite the extensive research in the area, an innate MC method with plenty of instances in nature, i.e., olfactory or odor communication, has not been studied with the tools of information and communication technologies (ICT) yet. Existing studies focus on digitizing this sense and developing actuators without inspecting the principles of odor-based information coding and MC, which significantly limits its application potential. Hence, there is a need to focus cross-disciplinary research efforts to reveal the fundamentals of this unconventional communication modality from an ICT perspective. The ways of natural odor MC in nature need to be anatomized and engineered for end-to-end communication among humans and human-made things to enable several multi-sense augmented reality technologies reinforced with olfactory senses for novel applications and solutions in the Internet of Everything (IoE). This paper introduces the concept of odor-based molecular communication (OMC) and provides a comprehensive examination of olfactory systems. It explores odor communication in nature, including aspects of odor information, channels, reception, spatial perception, and cognitive functions. Additionally, a comprehensive comparison of various communication systems sets the foundation for further investigation. By highlighting the unique characteristics, advantages, and potential applications of OMC through this comparative analysis, the paper lays the groundwork for exploring the modeling of an end-to-end OMC channel, considering the design of OMC transmitters and receivers, and developing innovative OMC techniques.
\end{abstract}


\section{Introduction}
\label{sec:intro}

\IEEEPARstart{L}{iving} organisms have evolved to detect various physical, chemical, and biological signals via their sensory systems. Among all, olfaction stands out as being highly pervasive, immediately transferred to the brain, and interconnected with the limbic system \cite{wang2018conceptual}. Especially, the human olfactory system directly interacts with the parts of the forebrain, which are responsible not only for odor discrimination but also for emotion, motivation, and certain memories. Processing the incoming odor-coded information within the limbic system results in a strong relationship between odors and psychological and cognitive processes.

\begin{figure}[t!]
	\centering
	\includegraphics[width=1\columnwidth]{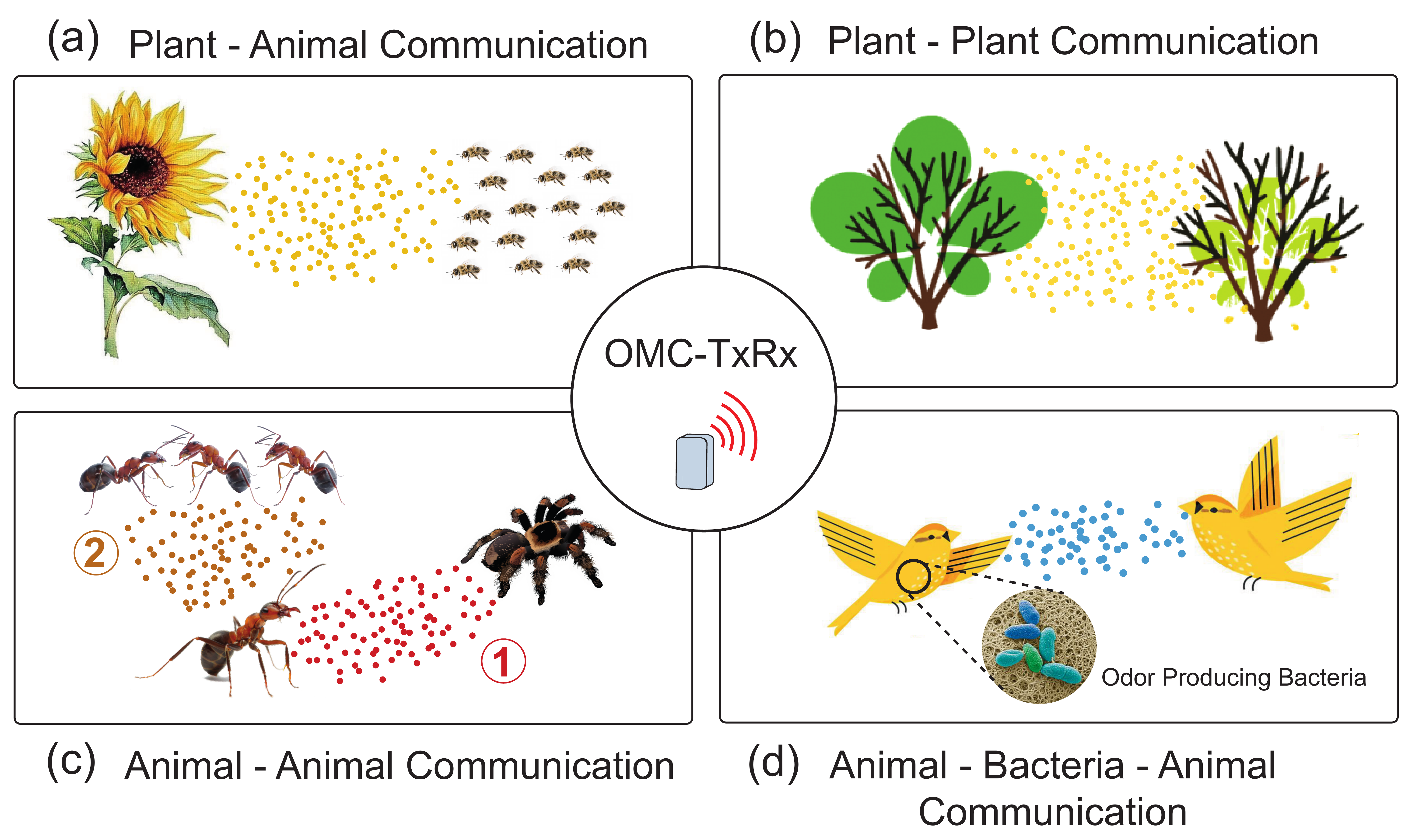}
    	\caption{Odor communications in nature:\\
(a) Plants use odor molecules to attract pollinators and to aid with seeding, \\
(b) Plants detect odor molecules to react to disease in nearby plants, communicate the presence of stressors, and coordinate behaviors, \\
(c) Animals detect incoming predators via odor molecules (1), release odor-based warning signals (2) within species, and may mimic a scent for exploitative reasons, \\
(d) Animals attract their mates by releasing pheromones produced by bacteria to recognize the identity of mates and offspring, for bonding, and for profiling state and intent.}
	\label{fig:mainFigure}
\end{figure} 

Thanks to its unique properties, olfactory communication enables interkingdom cross-talk between various organisms, as shown in Fig.~\ref{fig:mainFigure}. Immobile organisms, such as plants, use odor molecules to attract insects or animals to transfer their pollen to other plants or to detect diseases in nearby plants to protect themselves by eliciting defense responses. On the other hand, some animals use pheromones, i.e., odor-based vital signals, as a mode of communication for identifying individuals, indicating aggression/submissions, marking territories, and governing reproductive behaviors, i.e., to survive and continue their bloodline.

Understanding the fundamentals of natural olfactory signaling between different species and even kingdoms provides broad opportunities for controlling and interfering with these communication channels. Despite its huge potential, however, the principles of odor-based information transfer (from odor formation and coding to transmission, reception, and beyond) have not yet been thoroughly investigated, compared to those of other sensory systems, e.g., vision, audition, and somatosensation, from an information and communication technology (ICT) perspective.

With the help of ICT tools and techniques, revealing the fundamentals of olfactory communications can help realize the first artificial and controlled odor-based information transfer between humans and human-made things and interface the processes that the human olfactory system intertwines with. Especially due to its relation with the limbic system, deciphering this sense holds a huge potential for improving the mental and physical state as well as the performance of humans. Hence, this paper aims to review the state-of-the-art and outline what is needed to tame and engineer yet another natural phenomenon as a new communication modality in the Internet of Everything (IoE).

Uncovering the mechanisms of olfactory signaling will pave the way for unprecedented applications in the IoE domain, e.g., odor-based indoor and outdoor MC applications, early diagnosis of various diseases, such as cancer, diabetes, and a better understanding of chemical perception.

To guide the field in achieving this ambitious goal, this paper introduces the concept of olfactory communication by examining natural olfaction, odor information channels, reception mechanisms, spatial perception, and cognitive mapping. While recognizing that the terms odor, smell, and scent can be used interchangeably, with “smell" often associated with the sensory system and sensory experiences, and “scent" frequently used for pleasant-smelling odors, in this paper, we choose to consistently use the term “odor." Our primary focus is on particle-based communication modalities and exploring the use of odors as information molecules (IMs) in communication. By using odor throughout the paper, we emphasize the communicative aspect rather than solely focusing on sensory experiences. This choice allows us to highlight the broader implications of odors for cognitive and behavioral functions, going beyond their role in triggering sensory perceptions.

The paper starts by introducing the concept of odor-based communication and provides an overview of the current challenges, concerns, and future research directions in the field. Notably, it begins with a comprehensive comparison between various communication systems. The comparison includes molecular communication, natural olfaction, machine olfaction, and the specific focus of this paper: odor-based molecular communication, which is referred to as OMC from this point forward.

By examining these different communication systems, the paper highlights the unique characteristics, advantages, and potential applications of OMC. This comparative analysis sets the stage for exploring the modeling of an end-to-end OMC channel, the design considerations for OMC transmitters (Txs) and receivers (Rxs), as well as the development of innovative OMC techniques. Finally, concluding remarks are given.

\section{Odor Communication in Nature}
Communication is the process of transmitting and receiving information through a medium to develop a common understanding of transmitted messages. In nature, living organisms communicate by using verbal and nonverbal communication schemes. Nonverbal communication is the transmission of information between a transmitter and receiver without using words or language. Odor communication, also called olfactory communication or natural olfaction, is a nonverbal communication scheme that enables information transfer between living organisms using their sense of smell \cite{persaud2013neuromorphic}.

Olfactory systems evolved to detect incoming odor molecules, estimate their strengths, identify their source, and recognize a specific odor in the background of another. In the animal kingdom, two distinct olfactory systems have evolved for mammals and insects. For mammals, the olfactory system directly affects physiological regulation, emotional response, reproductive functions, e.g., sexual and maternal behaviors, and social behaviors, e.g., recognition of members of the same species, family, clan, or outsiders. For insects, the olfactory system involves modifying behaviors associated with mating, foraging, recognition of kin, brood care, swarming, alarm, and defense \cite{persaud2013neuromorphic}. 

Olfaction is one of the primary animal senses, and different animals utilize different olfactory organs. In the animal kingdom, organisms are divided into two categories: \textit{i)} vertebrates; \textit{ii)} invertebrates. The sense of smell is mediated by specialized sensory cells of the nasal cavity in vertebrates and, by analogy, the sensory cells of the antennae in invertebrates. Even though the olfactory systems evolved in various forms, the olfactory receptor neurons have common properties and structures. The existence of this shared foundation serves as the basis for a unified framework that can be used to examine how species communicate through odor molecules. By exploring the essential characteristics of olfaction, scientists can uncover the underlying mechanisms that govern odor perception and communication. This enables them to gain valuable insights into the universal principles that dictate how odor communication functions across all animal species.

Future developments in odor-based molecular communication (OMC), incorporating odor mixtures, specialized transmitter and receiver structures, and advancements in channel capabilities, hold significant potential for enabling inter-kingdom communication. Unlike classical molecular communication, OMC goes beyond the use of a single type of molecule and explores the transmission and reception of odor mixtures. The advent of OMC transceivers, as depicted in Fig.~\ref{fig:mainFigure}, opens up new avenues for exploring the intriguing realm of cross-species odor communication. This technological advancement not only offers the possibility of establishing communication channels between different species but also contributes significantly to our broader understanding of olfactory systems. OMC has the ability to unveil previously unattainable insights into the complex mechanisms underpinning inter-species odor communication by further expanding and utilizing its capabilities.

\begin{figure}[!t]
\centering
\includegraphics[width=0.9\linewidth]{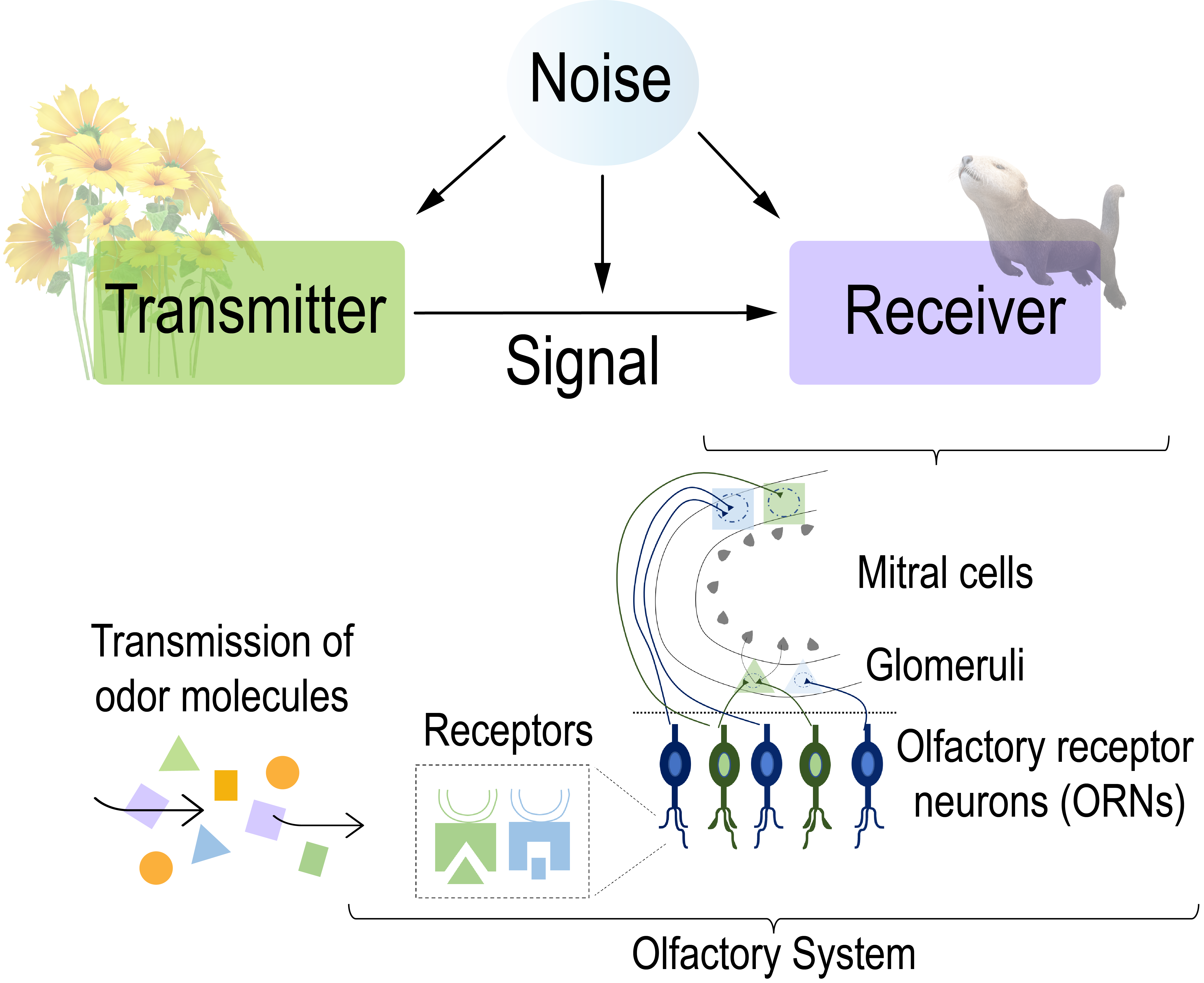}
\caption{Shannon's basic communication model (1948) extends to the interaction between plants and vertebrate animals. In this context, plants emit odor molecules as signals, which are detected and interpreted by the olfactory system of vertebrate animals, represented by their biological circuitry/transduction cascade. This analogy recognizes the possibility of noise interference in this communication process \cite{wilson2015noisy, shannon1948mathematical, Psynso}.}
\label{natural}
\end{figure}

\subsection{Odor Information, Channel \& Reception}
 
Odor-coded information is sensory data conveyed by odorous substances and is characterized by the electrical impulses produced by olfactory receptors in response to certain odor molecules \cite{persaud2013engineering}. Odor molecules possess distinctive features such as small size, volatility, and hydrophobicity \cite{qin2023artificial}. 
 
Odor molecules are commonly present in complex mixtures, and the human sense of smell can perceive these mixtures as distinct aromas. One approach to understanding odor perception is to attempt to identify and classify fundamental odor qualities that can account for a wide range of perceived odors. %Researchers have defined ten primary odors, including floral, woody, and fruity. 
A crucial understanding of the underlying neural mechanisms and the ability to distinguish various odors can be gained by classifying these main odors, including floral, woody, fruity, and others. Moreover, identifying primary odors can help develop tools for synthesizing or manipulating odors and aid in the design of odor-based products such as perfumes. %and air fresheners.

Nevertheless, the question persists as to whether it is possible to mix these primary odors to produce a full range of realistic real-world fragrances and to combine them to generate odors that are not only more realistic but also intricately complex~\cite{primaryodors}.

 \begin{figure}[!t]
\centering
\includegraphics[width=0.8\linewidth]{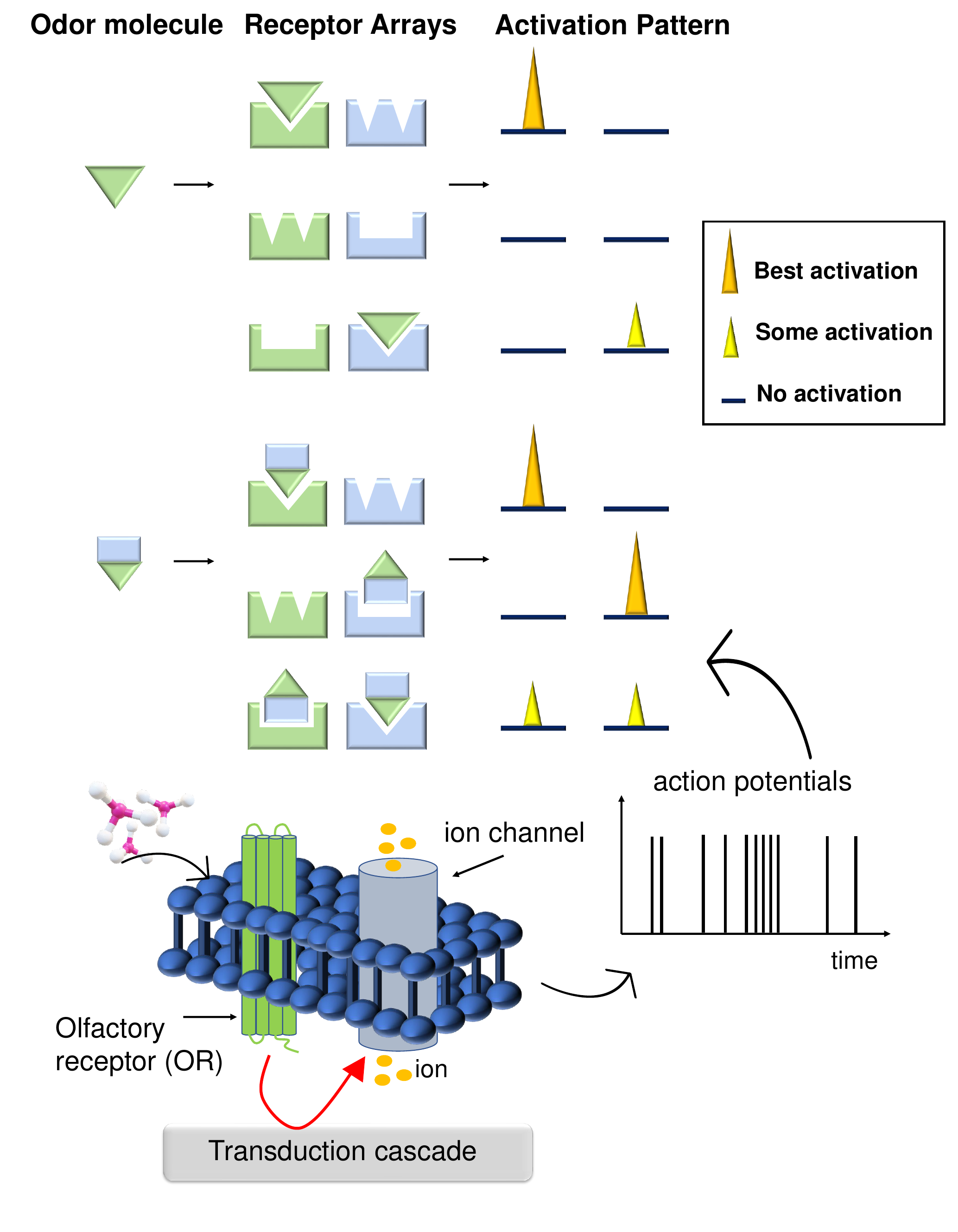}
\caption{Analogous to the generation of action potentials in neural processing, where sensory input triggers a transduction cascade and the generation of action potentials, the odor reception process is commonly modeled by the concepts of combinatorial binding and cross-reactive receptors \cite{manzini2022principles, buck2004unraveling}.}
\label{cb}
\end{figure}

Due to their small size and volatile nature, odor molecules can diffuse through the air and carry information for short distances without guidance, where guidance, such as airflow or heat flow, is required to reach longer distances.
In nature, the transmission of odor information occurs through the agency of wind or turbulent airflow. These natural guidance mechanisms enable the intended recipients to receive odor information at distant locations. 
The initial dispersion of odors is influenced by the starting point of odor release, which depends on variables such as position and strength of emission. The concentration of molecules decreases with distance, resulting in lower odor saturation due to factors like emission rate, space dimensions, and airflow patterns. Factors related to air quality, such as temperature and humidity, also play a role in the transmission of odors. Additionally, studying temporal patterns associated with scents offers valuable insights on odor communications \cite{patnaik2018information}. In Fig. \ref{natural}, Shannon’s basic communication model is used as an analogy to extend the interaction between plants and vertebrate animals \cite{wilson2015noisy}. In this scenario, information exchange takes place in a channel that is susceptible to noise. Vertebrate animals rely on their olfactory system, consisting of specialized biological circuitry, to detect and interpret odor molecules emitted by plants as signals. This process involves receptor neurons equipped with olfactory receptors that bind to specific odors, initiating a transduction cascade. The cascade triggers the opening of ion channels within the neuron, resulting in depolarization and potentially leading to action potentials being generated. Additionally, the olfactory system encompasses mitral cells and glomeruli located in the olfactory bulb, which contribute to further processing and interpretation of odor-related information \cite{manzini2022principles, persaud2013engineering,cheok2018digital}.

In odor communication, the reception process is commonly modeled through two important concepts: combinatorial binding \cite{malnic1999combinatorial} and cross-reactive receptors, i.e., the shape-pattern theory of olfactory perception \cite{jamali2023olfaction, buck2004unraveling, albert2000cross}. Combinatorial binding refers to the specific reactions or patterns of activation in the receptor array that occur when different combinations of molecules are present. This mechanism enables odor information to be encoded based on distinct patterns of receptor activation, analogous to how specific patterns of action potentials encode information in neural networks. This analogy highlights that different patterns of receptor activation correspond to different odor information, as illustrated in Figure \ref{cb}. On the other hand, cross-reactive receptors exhibit the ability to respond to multiple odors, expanding the range of detectable odor molecules. The receptor array is capable of being activated by multiple odors, albeit with varying degrees of activation. This cross-reactivity allows for the detection and discrimination of a wider variety of odors without requiring a dedicated receptor for each specific odor.

%Animals, in particular, have evolved to rely on these mechanisms to navigate their environments and locate food, mates, or predators. 

\subsection{Spatial Perception \& Cognitive Mapping}

Based on our current understanding, animals can extract spatial information, such as location, from the temporal patterns of odor signals carried by the wind. They can construct a cognitive map of their surroundings and make informed decisions about movement patterns by analyzing the changing concentration and composition of odor molecules, using the spatiotemporal information embedded into the received odor signals \cite{menzel2015memory, bao2019grid}. %However, the precise mechanisms underlying these abilities in an odorous landscape still need to be fully understood. 
Nevertheless, the intricate mechanisms that underlie these abilities within the context of an olfactory-rich environment are still awaiting comprehensive elucidation. Olfaction research often relies on confined spaces, such as rooms or laboratories. Although these controlled environments help researchers to examine odor communications, clearance mechanisms might be required in practical applications to remove persistent odor molecules that can linger in the air for longer than other signals, such as sound waves. 
 
Odor information can carry and trigger various types of sensory information. Categorizing odor information is an active research area and can be investigated under three categories: semantic, contextual, and quantitative. The semantic category refers to recognizing specific odors or odor mixtures, while the contextual category relates to the situation or environment in which the odor is perceived. The quantitative category pertains to the concentration or intensity of the odor stimulus. 

 Moreover, odor information have the ability to play a role in creating sensory information linked with spatial location. While the sense of smell does not directly offer accurate spatial coordinates, combining odor cues with other sensory experiences can improve our understanding of spatial context. This collaboration allows us to connect specific smells with particular places or surroundings, aiding in the formation of a cognitive spatial map \cite{Fischler-Ruiz2019Olfactory}. By connecting odor data to hints about space and location, we deepen our comprehension of how we perceive scents in relation to our environment and enhance our capacity for effective navigation and interpretation.

\begin{table*}[htbp]
  \centering
  \small
  \raggedleft
  \caption{Comparison of Different Communication Systems.}
  \label{tab:communication-systems}
  \renewcommand{\arraystretch}{1.2} % Adjust the vertical spacing
  \setlength{\arrayrulewidth}{1pt}
 \begin{tabular}{|p{2.4cm}|>{\arraybackslash}m{3.6cm}|>{\arraybackslash}m{2.8cm}|>{\arraybackslash}m{3.2cm}|!{\color{blue}\vrule width 1.2pt}>{\arraybackslash}m{3.5cm}|}
    \hline
    \textbf{\textit{Features}} & \makecell[l]{\textbf{Molecular}\\ \textbf{Communication (MC)}\\ \textbf{\cite{MALAK201219,phymc,suda2005exploratory,mcbody,mcfunda,mcict,mcnet,mcrev,hiyama2006molecular}}} & \makecell[l]{\textbf{Natural }\\ \textbf{Olfaction}\\ \textbf{\cite{wilson2015noisy, manzini2022principles,semin2019inter,ferkin2018odor,ezenwa2014microbes,jurgens2017changing,roberts2020human,carthey2018extended}}} & \makecell[l]{\textbf{Machine}\\ \textbf{Olfaction}\\ \textbf{\cite{covington2021artificial,ranasinghe2012digital, cheok2018digital, spence2017digitizing, gutierrez2014advances, kim2022artificial, pearce2006handbook, guthrie2017machine, marco2012signal}}} & \makecell[l]{\textbf{\textit{\textcolor{blue}{Odor-Based Molecular}}}\\ \textbf{\textit{\textcolor{blue}{Communication (OMC)}}}\\ \textbf{\cite{jamali2023olfaction, harel2003towards, mcguiness2018parameter, giannoukos2018chemical, giannoukos2017molecular, mcguiness2018asymmetrical}}} \\ \hline\hline
    \makecell[l]{\textbf{\textit{Information}}\\ \textbf{\textit{Carrier}}} & Synthetic pharmaceuticals, ions, nucleic acids, neurotransmitters, proteins \cite{kuscu2019transmitter, mcfunda} & Naturally occurring odor molecules, e.g., pheromones & Digitally encoded odor representations & Odor molecules and mixtures \\ \hline
    \textbf{\textit{Transmission Medium}} & Liquid or gas medium  & Liquid or gas medium & Liquid or gas medium & Liquid or gas medium \\ \hline
    \makecell[l]{\textbf{\textit{Encoding}}\\ \textbf{\textit{Mechanism}}} & Concentration, type, release time \cite{kuscu2019transmitter} & Modulation through chemical reactions & Digital representation of odors, e.g., olfactory displays \cite{kaye2004making} & Modulation of both concentration and types, i.e, mixture \\ \hline
    \textbf{\textit{Reception}} & Passive/Absorbing receiver, reactive receiver \cite{kuscu2019transmitter}  & Biological receptors, neural processing  & Chemical detectors and separators, E-Nose \cite{manzini2022principles, qin2023artificial}&  Cross-reactive receptor-based reception \cite{jamali2023olfaction}, chemical detectors \cite{mcguiness2019experimental} \\ \hline
  \textbf{\textit{Data rate}} & Relatively low data rate compared to EM communication, limited by molecule diffusion rate and environmental factors & Relatively low data rate compared to EM communication, {slower compared to vision due to complex odor mixtures \& biological mechanisms  \cite{szyszka2012speed}}  & Variable data rate depending on sensing technology, processing capabilities, and algorithms. Slower compared to vision, and faster compared to taste \cite{prasad2016human}  & Relatively low data rate compared to EM communication, influenced by molecule diffusion rate and complexity of odor mixtures\\ \hline
    \makecell[l]{\textbf{\textit{Practical Apps.}}\\\textbf{\textit{and Use Cases}}} & Drug delivery, lab-on-a-chip systems, nanonetworks, underwater communication & Interkingdom communication, ecological interactions  & Digital scent technology, fragrance industry, virtual reality & Chemical sensing, environmental monitoring, healthcare, security \\ \hline
  \end{tabular}%
\end{table*}%

Both animals and humans share commonalities in terms of odor reception. For example, fruit flies have been experimented with to understand and mimic the human olfactory system \cite{Stocker2001Drosophila, Liu2015Classical, Masek2013Drosophila}. As shown in Fig.~\ref{fig:mainFigure}, olfactory signaling is prevalent across different kingdoms, indicating a universal mechanism of chemical communication. Receptors in the olfactory system are sensitive to different metrics, such as the identity, intensity, and temporal pattern of odor molecules \cite{Spors2002Spatio-Temporal}. This information is gathered through sniffing, which involves the sampling of odor molecules that enter the nasal cavity. Once odor molecules bind to receptors, a combinatorial binding model, see Fig. \ref{cb}, is used to generate a unique code that is then transmitted to higher levels in the brain, where filtering, dimension reduction, and mapping are performed.

The olfactory system has an exclusive link with the brain. It has specialized pathways to the limbic system, which is strongly connected to emotions and memory \cite{Wenzel1968Olfactory, Bell1992An}. Olfaction plays a crucial part in various uses involving scents, such as driver focus, remembering smells, and personal experiences. Although the olfactory system's unconscious way of triggering processes can be advantageous, it may also introduce limitations since individual experiences and backgrounds can influence odor perception. Despite these challenges, a better understanding of the mechanisms underlying odor reception can lead to novel odor-based applications and improve our perception of olfaction in communication and cognition.

\section{ICT Understanding \& Engineering of Odor Communication}

Odor communication, an innate method found in nature, 
%has been relatively understudied despite its potential applications in areas such as navigation, foraging, and communication. The sense of smell 
plays a crucial role in how animals interact with their environment, including navigation, foraging for food, recognizing predators and prey, and communication with conspecifics. However, in contrast to other sensory modalities, we have a limited understanding of how olfaction can be used for communication, particularly in terms of end-to-end odor-based communication channels and principles of odor-based information coding. While research has focused on digitizing the sense of smell and developing actuators and sensors for olfactory displays and classification, a comprehensive understanding of end-to-end odor communication systems is still lacking. A critical step in harnessing the full potential of olfactory communication is to examine the fundamental principles that govern it from an information and communication-theoretical point of view. That includes understanding the encoding and decoding mechanisms of odor-based information, the structural and functional properties of the olfactory systems, and the natural ways of odor-based information transfer.

In Table \ref{tab:communication-systems}, we present a comparison of different communication systems, encompassing classical molecular communication, natural olfaction (natural odor communication), machine olfaction (digitizing sense of smell), and odor-based molecular communication (OMC). This table serves as a foundation for delving into the intertwined nature of these research domains in this paper. Our focus lies in understanding the ICT perspectives underlying natural odor communication while drawing inspiration from the advancements in machine olfaction, e.g., E-Nose technology, for the design and modeling of odor molecular communication (MC) transceivers. Additionally, we will discuss the existing MC techniques employed in classical molecular communication, using them as a starting point to develop innovative techniques tailored to the realm of advanced odor-based communication.

The objective of this paper is to lay the groundwork for future investigations into OMC by studying its occurrence in nature. Rather than solely focusing on identifying essential components like the spectrum, symbol, and channel, this study aims to contribute to a broader understanding of the fundamental aspects of odor communication. It provides insights and knowledge that can serve as a foundation for future investigations in this field. By exploring these fundamentals from an ICT perspective, this study aims to inspire further research and facilitate a comprehensive understanding of this communication modality. This understanding is crucial for unlocking the complexity and richness of this communication modality, enabling potential applications in areas such as developing olfactory interfaces for wireless connectivity and realizing the concept of the IoE \cite{IoE}. By comprehending the basic principles of how odor communication occurs in nature and how odors carry encoded information, we can develop effective olfactory interfaces and odor transceivers, leading to new applications in navigation, communication, and beyond.

%Olfactory communication, an innate MC method with plenty of instances in nature, has not been thoroughly studied. The existing efforts on odor research focus on digitizing the sense, developing actuators, and the odor classification based on a sensor array simulating the olfactory receptors (ORs) with a pattern recognition mechanism functioning like the olfactory bulb of humans, i.e., machine or artificial olfaction. However, end-to-end odor MC channels and the principles of odor-based information coding are not considered. Therefore, the ways of odor-based MC present in nature are need to be anatomised to unleash its application potential. Accordingly, the main constituents, e.g., spectrum, symbol, channel, in this uncharted territory, where the carrier is not an EM wave but information-carrying odor molecules can be revealed by studying the fundamentals of this unconventional communication modality from an information-theoretical point of view.

%Thereby incorporating the universe of odor molecules into the IoE framework, opening up numerous possibilities in the environmental, industrial and medical domains.

\begin{figure*}[t!]
	\centering
	\includegraphics[width=\textwidth]{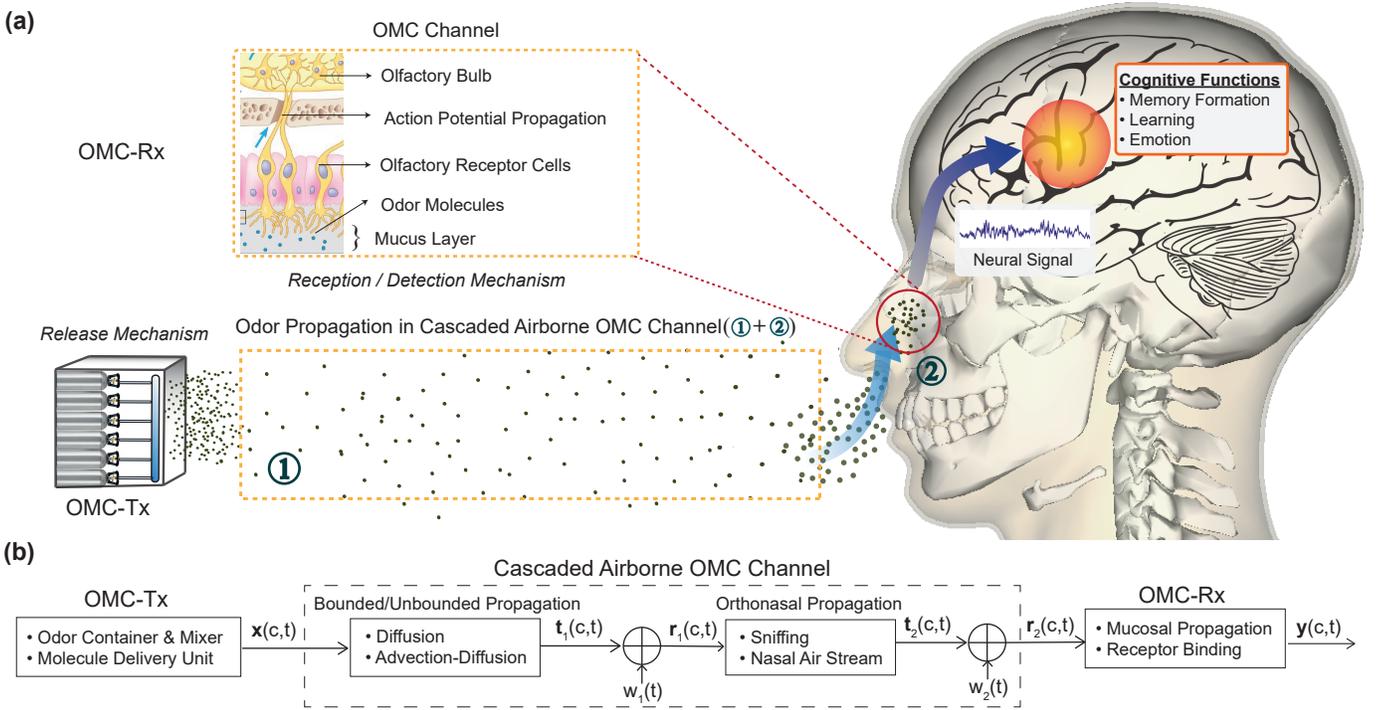}
    	\caption{End-to-end OMC system: (a) Holistic View, (b) System Components.}
	\label{fig:e2eMCsystem}
\end{figure*}

\subsection{Modeling and analysis of end-to-end OMC}

Humankind mimics the strategies that nature has perfected since its existence and uses them to address its problems. In this sense, the fundamentals of natural odor communication, e.g., pheromone communication in animals from an ICT perspective, need to be identified based on bio-inspiration, i.e., transferring and applying biological evolution and refinements for technological advancements. One of the major challenges is understanding how these natural solutions can be controlled, modified, or reengineered to transmit information that can be different from the natural. For example, the macroscale airborne MC systems developed use volatile chemicals, e.g., odors, as information molecules and adopt available technologies when implementing detectors, i.e., OMC-Rxs, while dispensers are used as OMC-Txs \cite{farsad2013tabletop, unterweger2018experimental}. Since the existing systems mostly consist of stand-alone devices, odor-based communication schemes and protocols need to be developed from a holistic perspective for an end-to-end MC system where information-carrying odor molecules/mixtures are sent and received from one point to another. To advance OMC, specific end-to-end channel models need to be developed, taking into account the non-linearity and temporal variance of OMC-TxRxs, as well as the unique properties of odor molecules. Although there exist theoretical work on MC, it cannot be directly applied to odor communication due to the distinct characteristics of odor molecules and the subjective nature of odor perception. Dedicated models are required to accurately capture odor transmission and reception, allowing for a deeper understanding and advancement of odor-based information transfer.

Modeling the process of odor transmission is essential for understanding how information is encoded and decoded using odors. Given that an OMC-Tx is placed on the outside of the nose, it is necessary to understand the cascaded airborne OMC channel, as illustrated in Fig. \ref{fig:e2eMCsystem}, for both bounded and unbounded medium, from the odor transmitter to the nose. By considering diffusion and advection-diffusion as propagation elements, we can better understand how odors travel and reach the nasal region. 

Additionally, it is important to model the nasal region itself, from the nose to the OMC-Rx, by taking into account the non-linearity and temporal variance of both the OMC transceivers and the channel conditions. These models should take into account the non-linearity and temporal variance exhibited by both the OMC transceivers and the channel conditions within the nasal region. These efforts in modeling contribute to a deeper comprehension of the mechanisms by which odors are detected and interpreted by the olfactory system. To accomplish this, it becomes essential to characterize the transmitted signal in terms of concentration profiles and the corresponding waveforms, based on the OMC transceiver-compatible channel geometry for airborne OMC channel. The type of application scenario, whether it is a bounded or unbounded transmission medium such as cylindrical geometry or \textit{tube}, will also be a crucial factor to consider. Overall, by developing a comprehensive and integrated approach for modeling the entire process of odor transmission, the potential for controlled information transfer using information-carrying odor molecules can be unlocked, whether it is for human or animal communication or other applications.

Orthonasal olfaction, specific to mammals, involves the detection of odors during sniffing and it is considered as a key mechanism for how mammals perceive their surroundings. However, other creatures, such as insects, birds, and even some mammals like dogs, rely on retronasal olfaction—sensing odors during exhalation—as their primary method for detecting odors, which highlights the diversity in the olfactory abilities among different species. In orthonasal olfaction, i.e., the detection of odors during sniffing, odor molecules enter the nasal region after traveling through the air stream and are captured by the ORs in the olfactory epithelium. There is a significant effort to accurately estimate airflow distribution patterns, i.e., nasal airflow patterns, while taking into account many physical variables such as initial concentration of odor molecules, inspiratory flow rate, and nasal anatomy. The airflow pathways are also investigated through the nose and characterize the airflow, e.g., laminar and mixed turbulent, depending on the nasal cavity structure, to predict airflow and odor delivery patterns. Computational models have also been developed in drug delivery research to simulate the transfer of drugs from the mucus layer to the olfactory bulb, focusing on nose-to-brain biopharmaceutics delivery \cite{kiparissides2019computational, pharmaceutics10030116}.

These extensive endeavors to model odor propagation within the nose can serve as the foundation for an integrated approach to comprehensively understand the mechanisms of olfactory sensation and to develop controlled information transfer using information-carrying odor molecules. By utilizing these models as building blocks, it is possible to aim for species-specific models or develop generic models by identifying dominant characteristics and similarities across different species. Understanding mucosal propagation is also crucial for modeling the end-to-end channel, making these models of both orthonasal and retronasal olfaction valuable for modeling the odor molecule communication receivers (OMC-Rxs) depicted in Fig. \ref{fig:e2eMCsystem}b). Even though sensors have been developed mimicking olfactory systems, they are suitable to operate in aqueous medium only. Therefore, a detection chamber has also been proposed to mimic the nasal cavity, utilizing gas permeable membrane-based systems integrated with odorant-binding proteins (OBPs), in order to replicate the mucus-like properties and provide an interface for gas detection \cite{choi2022bioelectrical}. Inspired by these approaches, the following sections will focus on modeling and designing pathways for OMC-Rxs, while also addressing possible challenges and providing potential solutions.

% Given that MC-Tx is placed on the outside of the nose, a cascaded airborne MC channel for the bounded/unbounded medium from MC-Tx to the nose, considering diffusion and advection-diffusion as propagation elements, and the nasal region, i.e., from the nose to MC-Rx, needs to be modeled by incorporating the non-linearity and temporal variance of MC-TxRxs and channel conditions. To this end, the transmitted signal in terms of a concentration profile and the corresponding waveform should be characterized based on the MC-TxRxs-compatible channel geometry for airborne MC. The application scenarios will also be decisive on this, e.g., bounded/unbounded transmission media, e.g., cylindrical geometry (\textit{or} tube). 

\begin{figure*}[!t]
\centering
\includegraphics[width=0.9\linewidth]{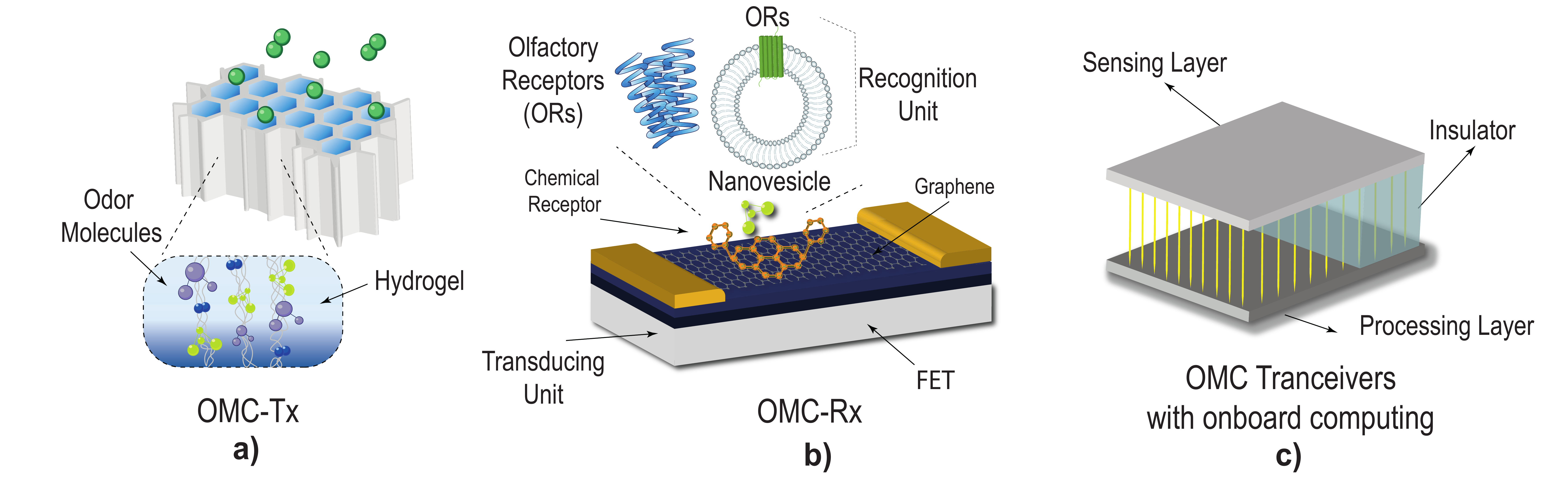}
\caption{Future OMC transceivers: a) Hydrogel-based OMC-Tx \cite{guo2023strong}, b) Graphene FET-based OMC-Rx with recognition layer comprised by either chemical or biological receptors \cite{tisch_haick_2010}, c) OMC tranceivers with onboard computing capabilities \cite{fan2023monolithic}.}
\label{txrx}
\end{figure*}

Challenges in modeling the end-to-end OMC channel include the varying diffusion coefficients of odor molecules, advective flow effects, and transmission distance. These factors affect the concentration profile of odorants, requiring accurate modeling to capture their complex interactions and ensure realistic odor transmission simulations. Moreover, from an ICT perspective, the characterization and modeling of the channel noise considering the physical channel properties, e.g., varying boundary conditions, obstacles, temperature fluctuations, and molecular crowding, is another major challenge. The characterization of the received signal and the reception noise should also be realized based on the odor entry mechanism, i.e., entry of odors from air to the nasal region where sampling of odor information is performed, their propagation in mucus \cite{koca2021molecular} considering the odor solubility in mucus, and the capturing of odors by ORs based on a specified binding model, e.g., combinatorial or competitive binding \cite{sharma2019sense}. The primary communication properties and functionality of the OMC system based on ICT metrics, such as channel capacity, bandwidth, delay, and error rates, are required to be analyzed to reveal its fundamental capabilities and limitations with the validation of the analytical channel models. The identification of these attributes will thus enable the development of different strategies to realize OMC for various IoE application scenarios, as well as the optimization of their implementation and design.

\subsection{Design and Modeling of OMC Transceivers}

Another major challenge in studying end-to-end OMC is devising compatible OMC transceivers. Even though there have been many studies on developing odor-based sensors that can serve as inspiration, conventional discrete and tuneable gas/chemical sensors, odor displays, and other components such as immobilized biological odorant-related proteins \cite{covington2021artificial} must be modified and integrated with the existing OMC transmitter and receiver models and designs. To achieve this, fundamentally new design and modeling methodologies for OMC-TxRxs must be obtained to govern the inherent capabilities, sensitivities, and fundamental limits of natural odor transmitting and receiving entities and to address the many difficulties that an odor communications system could potentially encounter. Therefore, from the engineering design perspective, it is essential to obtain the stochastic and analytical release model of an OMC-Tx, consisting of an odor container and mixer and a molecule delivery unit, regarding how many odor molecules it can emit and the range of odors available. Moreover, suitable theoretical shapes, e.g., point- or ring-shaped Tx, spherical or ring-shaped Rx, should be considered in the model for OMC-TxRxs tailored to the OMC system. The OMC-Rx design will pave the way for revealing the reception mechanism of the human nose and its relation with the limbic system, affecting psychological and cognitive functions. Thus, the designed OMC-Tx can be utilized to trigger specific functions via encoded odor molecules, referring to end-to-end controlled information transfer with odors. The OMC-Rx can also be implemented as a body implant to treat temporary or permanent smell loss, a common intra- or post-COVID implication. % Alternatively, it can be attached to human-made things as an artificial OR to use odors as the newest communication modality of the digital world, allowing machines and humans to communicate seamlessly in the concept of IoE.

%%This is quite a long section so it might help with clarity to break it down a bit and emphasise the significance in the titles

\section{Odor Signal Generators and Transmitters}

The current position of odor communication within the MC literature, particularly in macro-scale MC, involves the use of odor molecules as information carriers. Automated sprayers are commonly employed \cite{farsad2013tabletop, kim2015universal, exp19}; however, they typically possess a single storage area and lack the capability to generate odor mixtures. Alternatively, in-house-built odor generators serve as chemical pulse generators, but they often require bulky chambers and pipes \cite{mcguiness2018parameter,mcguiness2019experimental, giannoukos2017molecular, giannoukos2018chemical, mcguiness2018asymmetrical}. 

Olfactory displays, which involve emitting odor molecules or simulating the sense of smell through techniques like electro-stimulation \cite{patnaik2018information}, are an important inspiration and reference for designing physical OMC systems.  Although the main goal of this technology is to improve sensory experiences by combining scents with other senses like vision and sound, it also plays a role in designing, modeling, and transmitting artificial smells for OMC-Tx. Olfactory displays find various applications in fields including virtual reality and augmented reality, entertainment and gaming industries, and marketing campaigns \cite{Niedenthal2019A}.

When designing OMC transmitters, several design criteria from classical MC transmitters remain applicable \cite{kuscu2019transmitter}. Miniaturization is crucial, aiming to create compact and portable devices that can be seamlessly integrated into various applications. Molecule reservoirs should be designed to securely store odors, enabling controlled and precise release when required, as denoted in Fig. ~\ref{fig:e2eMCsystem}. This involves the implementation of both analytical and physically realized release mechanisms. Analytical methods, such as mathematical models and algorithms, can be employed to determine the optimal timing and quantity of odor release based on the communication requirements. Physically realized mechanisms, on the other hand, involve the design of reservoir structures and materials that can effectively contain the odors and release them in a controlled manner, for example, through diffusion, heat, or active pumping mechanisms \cite{kuscu2019transmitter}. By combining analytical and physical approaches, odor MC transmitters can achieve accurate and tailored odor release, enabling efficient and reliable communication.

Biocompatibility is especially important when considering intrabody environments where olfactory communication systems may be used. Ensuring compatibility with the human body and minimizing any potential adverse effects is essential for safe and effective operation. Transmission performance parameters such as OFF-state leakage, transmission rate precision, resolution, and range should be optimized to achieve reliable and accurate odor transmission \cite{kuscu2019transmitter}. Other design criteria include ensuring a reliable power source for continuous operation and designing compact and portable devices for ease of use and deployment, which implies on-board computing to be able to perform source/channel coding and modulation tasks successfully.

Indeed, one of the main distinctions between classical MC-Tx and OMC-Tx is the ability to store and mix multiple odors. OMC-Tx requires the capability to handle and release a diverse range of odor molecules, allowing for the encoding and transmission of complex odor information. This introduces an additional layer of complexity compared to classical MC-Tx systems, which typically involve the transmission of a single type of molecule.

Considering potential advancements and future directions can provide inspiration and guidance for design and modeling methodologies. Exploring speculative ideas for the future of olfactory communication systems allows us to envision captivating possibilities that might shape the way we encode and transmit odor information. Future design aspects could involve advancements such as:
\begin{itemize}
\item \textit{Diversified form factors}: Investigate a range of novel shapes for transmitters, encompassing designs beyond traditional forms. Experimenting with unconventional shapes, such as microfluidic arrays or flexible membranes, can offer opportunities to optimize odor transmission and reproduction in olfactory communication systems. By exploring innovative geometries, new possibilities can be unlocked for enhancing the efficiency and effectiveness of odor encoding and transmission \cite{covington2021artificial}. For example, the use of nanomaterial-based odor MC transmitters, such as hydrogel and graphene oxide films-based odor MC transmitters \cite{civastx}, represents a promising avenue for next-generation designs \cite{civas2023graphene, civas2022molecular} as illustrated in Fig.~\ref{txrx}a). These transmitters can store multiple odor molecules, enabling efficient and versatile odor transmission capabilities \cite{guo2023strong, nguyen2017biocompatible}.
\item \textit{Advanced release models}: Develop models that capture the stochastic nature of odor molecule emissions, enabling a deeper understanding and more accurate modeling of odor transmission.
\item \textit{Artificial odors}: Artificial odor research focuses on developing synthetic fragrances that closely mimic natural odors. The goal is to enable reliable communication by designing devices that can generate and emit specific odors in a repeatable and controllable manner. This ensures accurate encoding of odor information in odor molecular communication systems. Consequently, at the destination location, the replicated odors can be achieved through the use of artificial odors or alternative technologies capable of recreating the original odors.
\end{itemize}

By considering these design criteria and exploring future advancements, odor MC transmitters can be further optimized, opening up new possibilities for efficient and reliable odor communication systems.

\section{Odor Detectors and Receivers}

\begin{table*}[tbp]
\centering
\caption{Comparison of MC Receiver Architectures for OMC-Rx.}
\label{tab:MCRxs}
\begin{tabular}{p{3.5cm}p{3.5cm}p{5cm}p{4cm}}
\toprule
\textbf{Receiver Architecture} & \textbf{Limitations as MC-Rx} & \textbf{Suitability for OMC-Rx} & \textbf{Adaptation to OMC-Rx} \\
\midrule
\multicolumn{4}{l}{\textbf{Macro-scale Approaches}} \\
\midrule

\makecell[l]{\textit{\textbf{Magnetic Nanoparticle}}\\ \textbf{\textit{Detector } \cite{bartunik2019novel}} } 
& Bulky coils, limited selectivity & Adaptation required for broad-spectrum IM detection & Suitable for customization to specific OMC applications such as simple odor classification and concentration-based detection. \\
\makecell[l]{\textit{\textbf{Mass}}\\ \textbf{\textit{Spectroscopy} \cite{mcguiness2018experimental, mcguiness2019experimental, mcguiness2019modulation}}} & Bulky equipment, limited scalability  & Adaptation required for broad-spectrum IM detection & Needs simplification and automation \\
\makecell[l]{\textbf{Gas Sensor-based}\\ \textbf{\textit{MC-Rxs} \cite{farsad2013tabletop, koo2016molecular}}} &  Regeneration and saturation problem, limitations in sensitivity and accuracy & Limited suitability, improvement in signal processing and calibration required & Additional sensor types needed for diverse compound detection\\
\midrule
\multicolumn{4}{l}{\textbf{{Micro/nano-scale Approaches}}} \\
\midrule
\makecell[l]{\textit{\textbf{Nanomaterial-based }}\\ \textbf{\textit{FET MC-Rxs} \cite{sinw,kuscu2021fabrication, kuscu2016physical}}} & Cannot detect neutral IMs, prone to screening effect, lacking a precise design of recognition layer for different kinds of molecules to be recognized  & Not directly applicable to odor mixtures, specialized recognition unit required. Operates in aqueous solution, posing challenges in detecting insoluble  odors in water & Adaptation of recognition unit for odor mixtures required and/or sensor array. Additional interface required for gas detection, e.g., gas permeable membrane \cite{choi2022bioelectrical} \\
\makecell[l]{\textit{\textbf{Mechanical}}\\ \textbf{\textit{MC-Rxs}  \cite{aktas2022weight, aktasd}}} & Lacking a precise design of recognition layer for different kinds of molecules to be recognized & Most suitable by being able to apply type and concentration-based modulation \& detection, robust to ISI, higher capacity, increased operating range
 & Adaptation of recognition unit for odor mixtures required and/or sensor array \\
\bottomrule
\end{tabular}
\end{table*}

In the field of molecular communication (MC) literature, odor communication is currently addressed using various odor detectors and receivers. Basic gaseous sensors \cite{farsad2013tabletop, kim2015universal} and more sensitive detectors like mass spectrometers \cite{mcguiness2018parameter, mcguiness2019experimental} are commonly used. These devices, also known as portable trace detection systems, analyze the chemical composition of odorous substances using techniques such as gas chromatography, ion mobility spectrometry, or mass spectrometry. While gas chromatography-mass spectrometry technologies offer excellent specificity in identifying individual volatile components, they are often bulky and lack real-time detection capabilities.

Table \ref{tab:MCRxs} presents a comparison of various MC-Rx architectures for OMC-Rx adaptation. While there have been efforts to design MC-Rx receivers based on synthetic biology \cite{unluturk2015genetically, nakano2014molecular}, these systems have limited computational capabilities, can only operate with \textit{in vivo applications}, and are incompatible to interface cyber networks, despite offering high biocompatibility \cite{sinw}. In order to address the complexity of odor mixtures and signals, it is crucial to have odor detectors and receivers with advanced detection and processing capabilities. Furthermore, to ensure universal designs and scalability, it is necessary for these detectors and receivers to operate effectively in both in vitro and in vivo environments. As a result, we include artificial MC-Rxs in the table for comparison, considering their potential to meet these requirements and provide universal applicability. The macro-scale approaches listed in the table include the magnetic nanoparticle detector \cite{bartunik2019novel}, mass spectrometry \cite{mcguiness2018experimental, mcguiness2019experimental, mcguiness2019modulation}, and gas sensor-based MC-Rxs \cite{farsad2013tabletop, koo2016molecular}. These approaches have limitations such as bulky equipment, limited selectivity, and issues with regeneration and saturation. They need adaptation for broad-spectrum IM detection and may require improvements in signal processing and calibration.

On the micro/nano scale, there are two distinct approaches to consider. The first approach is nanomaterial-based Field-Effect Transistor (FET) MC-Rxs \cite{sinw, kuscu2021fabrication, kuscu2016physical}, which include designs utilizing graphene and silicon nanowire (SiNW). However, these nanomaterial-based FET MC-Rxs have limitations. They cannot detect neutral information molecules, are prone to the screening effect, and lack a precise design of the recognition layer for different types of molecules to be recognized. As a result, they are not directly applicable to odor mixtures, and adaptation of the recognition unit and potentially a sensor array are required to overcome these limitations.

On the other hand, the second approach, Mechanical MC-Rxs \cite{aktas2022weight, aktasd}, holds promise for odor detection. Mechanical MC-Rxs can apply type- and concentration-based modulation and detection, making them highly suitable for odor recognition. They exhibit robustness against intersymbol interference, have a higher capacity, and operate over a wider range. However, in the context of odor mixture recognition, adaptation of the recognition unit specifically designed for odor mixtures, along with the potential utilization of a sensor array, are still necessary.

\begin{figure*}[!t]
\centering
\includegraphics[width=\linewidth]{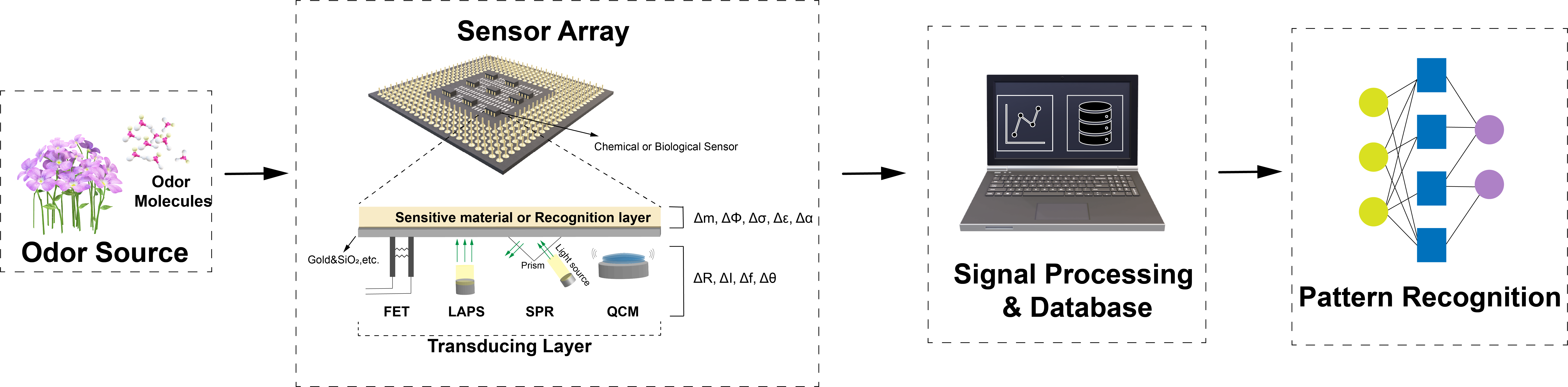}
\caption{Illustration of the E-Nose concept \cite{qin2023artificial}.}
\label{enose}
\end{figure*}

In addition to existing MC-Rx designs, E-Nose technology offers a promising avenue for adaptation as OMC-Rxs. E-Nose systems have the potential to fulfill the criteria for a functional OMC receiver, including in situ, continuous, and label-free operation \cite{kuscu2016physical}. These systems consist of sensor components that emulate the sense of smell found in living organisms. By analyzing the signals from these sensors and applying pattern recognition and classification techniques, E-Nose technology can effectively decipher odor information, as illustrated in Fig. \ref{enose}. These sensors can be made up of various types like chemical arrays such as metal oxide arrays, surface acoustic wave sensors, or nanosensors \cite{khan2020nanosensor, li2014recent}.

Research has been conducted to explore the potential of using biosensor-based electronic noses, which replace chemical receptors with biological receptors. To enable continuous monitoring over an extended period of time, nanotechnological solutions have emerged. For example, carbon nanotube (CNT) transistors are utilized as transducers, and immobilized odor receptors serve as recognition elements \cite{goldsmith2011biomimetic}. Another development in this field is the creation of a nanovesicle-based E-Nose \cite{jin2012nanovesicle}. These advancements have greatly contributed to the progress and application of electronic noses in odor communication.

The OMC-Rx detects and interprets received odor signals. Design and modeling methodologies for receivers could involve the use of chemical sensors, olfactory interfaces, or E-Nose devices. Developing models that capture the binding and transduction mechanisms involved in the reception process will be crucial for understanding how the receiver interprets odor signals. This may require the use of sensors or technologies that are sensitive to specific chemicals or compounds. Suitable processing units will be necessary to interface the receiver outputs with digital systems, enabling further processing, data analysis, and integration with other communication modalities.

Fig. \ref{txrx}b) depicts a futuristic design for an OMC receiver based on graphene FET transducing. In addition to chemical receptors or sensitive materials \cite{manzini2022principles}, biological receptors such as odorant binding proteins (OBPs), ORs, and ORs embedded in nanovesicles can be immobilized as recognition units \cite{tisch_haick_2010, gao2021artificial}. Various transducing schemes like light-addressable potentiometric sensor (LAPS), surface plasmon resonance (SPR), and QCM can also be used \cite{qin2023artificial}.

The key distinction between current E-Nose devices and OMC receivers is that the latter are designed to work harmoniously with transmitter devices in a holistic end-to-end system. In intrabody health applications, OMC receivers must prioritize biocompatibility, continuous monitoring, and low power consumption. Stand-alone devices for OMC receivers are envisioned, capable of onboard computing as illustrated in Fig. \ref{txrx}c). In contrast, most E-Nose systems separate the sensors from the processing units, relying on external computers for data analysis. Therefore, future OMC transceivers require the development of in-sensor computing using two-dimensional materials with in-memory and transistor-based computing technologies \cite{shulaker2017three,fan2023monolithic,zhou2020near}. Furthermore, hybrid and integrated OMC transceiver structures, combining both transmitting and receiving functionalities within a single device, need to be developed, such as a bioelectronic nose combined with a microfluidic system \cite{lee2015bioelectronic, rebordao2020microfluidics}.

Another important aspect of OMC receiver design is the incorporation of olfactory interfaces. Olfactory interfaces play a crucial role in facilitating interaction between the olfactory communication system and users or other digital systems. Olfactory interfaces act as a bridge, seamlessly integrating olfactory communication with other technologies and enhancing the user experience. User-friendly interfaces unlock the full potential of olfactory communication, enabling meaningful interactions between humans and technology. This integration facilitates smooth communication between olfactory systems and technologies like virtual reality, augmented reality, and smart devices.

%% we could merge the above with these below sections

\subsection{Limitations of Current Odor Receiver Technologies}
Mass spectroscopy (MS) fails as a molecular identification tool for increasingly large molecules since molecular mass and charge are not unique properties; consequently, uncertainty scales as these do.
In an attempt to overcome this issue, E-Noses consisting of a combination of a mass spectrometer and Field Asymmetric Ion Mobility Spectrometry (FAIMS), which has the ability to differentiate shape, have been commercialized and are now considered the ‘gold standard'. The flaw in these devices is that they require a supply of inert gas, a complex and expensive fabrication procedure. Consequently, industry attempts to miniaturize this and improve portability have failed. Additionally, the high electric field and heating required can cause changes to the compound of interest. The sensitivity is limited by a quantity significant enough to cause a discernible peak in the spectra.
Organic semiconductor devices form the majority of E-Nose technologies, followed by MOS and then conductive polymers. All three technologies experience limited lifetimes due to fouling and drift, and the former requires protective strategies to mitigate against this. MOS suffers from limited sensitivity due to surface charge accumulation.
What these established technologies have in common with each other and the plethora of other devices that fall under the umbrella of E-Nose technologies is that they are functionalized. As a result, there is a lack of solutions able to map the profile of compounds in an exploratory manner in unpredictable environments. This would be achievable in a sensor that measures a property or properties of the molecule and distinguishes responses using data processing methods \cite{enose2022polyimide}. This is the advantage presented by the current ‘gold standard’ for ‘E-Nose’ devices discussed above, FAIMS in combination with MS.
It becomes clear that there is demand for E-Nose technologies relying on alternative methods of interacting with both the environment and the sensor. Specifically, there is 
need for E-Nose technologies capable of measuring 
\begin{itemize}
    \item Large molecules, which includes many compounds of biological origin,
    \item Accurately recognizing ratios of compounds, e.g., the ratio of compounds in human breath in a person not experiencing inflammation from any cause \cite{Smolinska2014}, 
    \item Simultaneously measuring multiple compounds for situations where the true picture can only be conveyed with the presence of different compounds, such as bacterial strain identification \cite{ling2023},
    \item Identifying the whole molecular shape accurately has advantages analogous to determining the information contained within a strand of RNA. Such advances would vastly extend the vectors through which information may be stored, e.g., via chirality.
\end{itemize}

\subsection{Advancing OMC through Innovations in E-Nose or Gas Sensor Technologies}
With the advent of new odor receiver technologies that can outperform current E-Noses by recognizing larger or complete molecules, significant advancements can be made in several areas. The subsequent developments may involve the following steps:
\begin{itemize}
\item With the newfound ability to respond to large biological molecules, the channels over which communication can take place will be extended, and the difficulty of interception will increase.
\item Advances in the ability to identify larger biological molecules may enable reading or research into biological communication processes and facilitate attempts at biomimicry and interception.
\item Intangent, artificial odors could be developed that may be complementary to the sensor and environment. These may be used to optimize sensitivity by balancing sensor response with diffusivity and minimizing noise.
Complex profiles of compounds may be explored for the significance of combinations and ratios.
\end{itemize}

\begin{figure*}[!t]
\centering
\includegraphics[width=\linewidth]{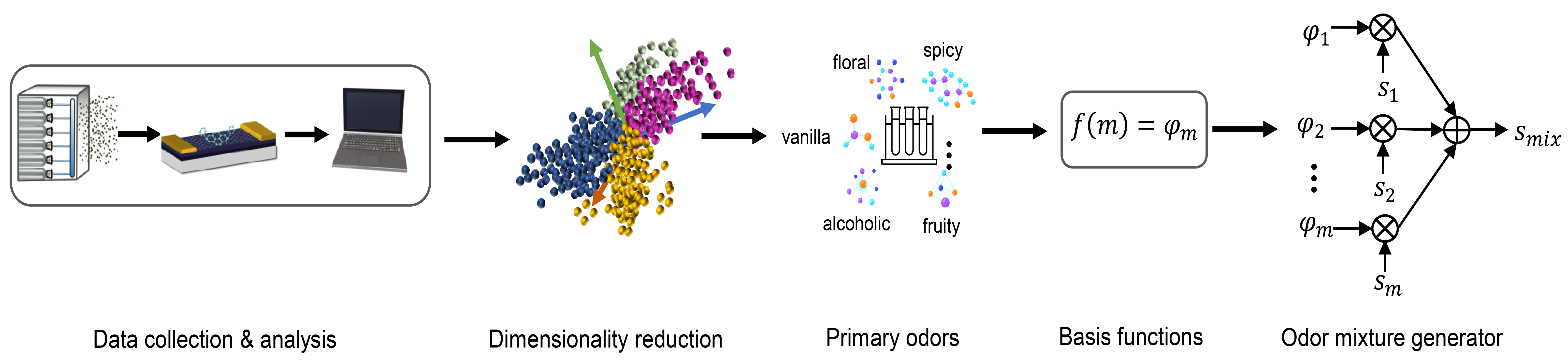}
\caption{Odor mixture generation based on primary odors.}
\label{prim}
\end{figure*}

\section{OMC Techniques}

In order to fully comprehend and implement an end-to-end OMC system, it is imperative to delve deeper into the field of communication techniques. Specifically, a comprehensive exploration of the fundamental traits of the human olfactory system is necessary in order to shed light on various unexplored aspects. By doing so, it is possible to effectively engineer effective modes of communication between humans and man-made objects while simultaneously addressing numerous challenges present within the existing literature on molecular communications.

%The realization of an end-to-end odor MC system requires the development of communication techniques based on the key characteristics of the human olfactory system to uncover many unknowns about it, thus enabling its engineering for communications between humans and human-made things while addressing many problems existing in the MC literature.  

\subsection{Modulation}

Upon sniffing, objective measures, i.e., qualitative classification of an odor and estimation of its strength, are performed by the brain \cite{harel2003towards}. Since molecule type and concentration are evaluated according to these objective measures in odor perception, advanced odor-based modulation can be developed based on the conventional modulation schemes tailored to MC that encode information based on many properties of the information-bearing molecules, e.g., concentration, type, and release time. Modulation schemes relying on combinations of different molecules will enhance the range and throughput of the MC system compared to conventional modulation schemes, e.g., molecule shift keying modulation. On the other hand, in mixture-based modulation schemes, the encoding mechanism involves the simultaneous modulation of both the concentration and type of odor molecules, as denoted in Table \ref{tab:communication-systems} \cite{jamali2023olfaction}. This approach broadens the capabilities and potential applications of molecular communication systems, offering increased range, throughput, and flexibility in encoding information using odor-based signals.

\subsection{Odor Waveform Generation}

In order to modulate odor signals, it is necessary to have an analytical framework that can generate different patterns of odor waveforms in a controlled manner. There are various approaches for classifying odors. One approach is based on empirical classification, also called odor profiling, where individuals classify odors based on their feelings or experiences, also called the hedonic valence-based approach. Another approach involves primary odor-based classification, which uses a reference set of known odors for categorization. Additionally, statistical techniques can be applied to analyze large amounts of olfactory data and extract meaningful classifications \cite{ghinea2011olfaction, chastrette2002classification}. These techniques can be used to identify and obtain primary odors, which serve as the basis for generating odor signal waveforms. 

As shown in Fig. \ref{prim}, OMC transceivers can allow for the acquisition of odor data sets, which can then be processed using dimensionality reduction methods such as principal component analysis (PCA) or independent component analysis (ICA) \cite{lee2022principal, harel2003towards}. These methods help in reducing the high-dimensional odor data into meaningful and interpretable lower-dimensional representations. However, to effectively analyze and interpret the acquired data, comprehensive information about odor-related components is essential.

Significant efforts have been dedicated to the creation of databases and olfaction wheels, which provide comprehensive information on various odor-related components. These include odors, odorants, and odorless compounds, along with their physicochemical properties. Furthermore, detailed data on olfactory receptors (ORs), odorant and pheromone binding proteins, and the interactions between ORs and odor molecules are being compiled, such as OlfactionBase \cite{sharma2022olfactionbase}. Access to these databases enables researchers to make informed decisions about odor classification and quantification based on physicochemical features. 

To effectively utilize the acquired data and information from databases, extensive efforts have been made to develop data pretreatment and feature extraction algorithms for odor classification and quantification. These algorithms help in extracting meaningful features from raw odor data and enhancing the performance of E-Nose systems. Additionally, machine learning techniques tailored to E-Nose systems have been developed \cite{ye2021recent, zarra2020instrumental, yan2015electronic, xu2021determination, spike, label, aguilera2012electronic, zhang2012classification}. Investigating the dimensionality of the odor space in physical, neural, and perceptual contexts has become a current focus, aiming to improve our understanding of odor discrimination \cite{meister2015dimensionality}. Similar to color vision, where three primary lights can adequately represent the perception of color, and just as audition utilizes sound pressure, frequencies, and variations as observable quantities, the chemical senses of olfaction and taste also require an equivalent set of observables for comprehensive understanding and practical applications. However, olfaction currently lacks this equivalent set of observables \cite{manzini2022principles}. Consequently, defining primary odors and establishing a vectorial odor space become crucial in bridging the gap between odor classification and ICT. By drawing parallels with signal generation in electromagnetic communication (e.g., frequency and amplitude) and image processing (e.g., pixels), this framework holds promise for integrating odor communication into ICT. Hence, by assigning these primary odors as orthonormal basis functions, denoted as $f(m) = \phi_m$, the combined odor mixture, $S_{mix}$, can be represented as a linear combination of these basis functions. This approach, as given in Fig. \ref{prim}, involves coefficients $S_1, S_2, \ldots, S_m$ that correspond to each basis function. Various factors such as concentration, molecule type, volatility, molecule structure, and distance between molecules in the mixture can be integrated into this concept. This concept enables the analysis and synthesis of odor mixture generation in an analytical fashion. The following metrics need to be considered when designing these concepts:

\begin{itemize}
    \item \textit{Comprehensive set of primary odors:} It is essential to incorporate a diverse range of primary odors that cover various scent categories. This comprehensive set ensures that the vectorial odor space can effectively represent different types of smells.
    
    \item \textit{Olfactory white:} Similar to visual or auditory whites, the concept of olfactory white refers to a neutral smell that converges towards a single olfactory note. In the design of the system, it is important to include olfactory white as a parameter. Olfactory white typically consists of mixtures of $\sim\,$30 or more equal-intensity components \cite{Weiss2012Perceptual}. 
\end{itemize}

Through this framework, the development of analytical methods and techniques can facilitate odor communication. By defining a vectorial odor space, it becomes possible to transform and manipulate odor signals in a structured manner, mapping input odors to desired output responses. This analytical framework offers valuable insights into the input-output relationship of odor communication from an ICT perspective. Various factors, including concentration, odor type, molecule structure, distance between molecules in mixtures, and volatility, can be taken into account when determining basis functions or vectors that accurately represent the properties of odor signals. This approach has the potential to pave the way for the establishment of controlled, standardized, and adaptable odor-based communication protocols.

\subsection{Odor Coding}
To engineer odor-based MC odor coding techniques, it is crucial to consider other fundamental properties of olfaction. One such property is the inhibition caused by specific molecules, which activate only certain types of receptors while inhibiting others. This aspect has been explored in the context of OMC \cite{jamali2023olfaction}. Additionally, the peripheral level of odor interactions and coding, which involves the modification of odor responses in the olfactory epithelium before capture, should also be taken into account. This includes investigating the potential interactions between odor mixtures within or between odor mixtures, which may lead to the formation of new chemical structures and modifications of odor binding sites \cite{kurian2021odor}.

%Other fundamental properties of olfaction, e.g., inhibition by some specific molecules activating only some types of receptors while inhibiting others, should be considered to engineer odor-based MC odor coding techniques as in \cite{jamali2023olfaction}.
% The peripheral level of odor interactions and coding, i.e., the initial level of modification of odorant responses in the olfactory epithelium, should also be accounted for.
%Odor interactions and coding in the initial level, i.e., modifications at the olfactory epithelium before capture, should also be accounted for. In this regard, it is important to investigate the possibility that odorant mixture interactions within or between odor mixtures might result in the formation of new chemical structures and the modification of odorant binding sites \cite{kurian2021odor}.

Moreover, pulse shaping for mixture-based modulations to improve the robustness of the receiver to intersymbol interference (ISI) can be developed by forming the transmitted odor mixture based on the physical and chemical characteristics of odor molecules, such as molecule size and volatility \cite{wicke2022pulse}. Different odor components in a mixture will experience different attenuation rates during propagation due to their unique physical and chemical properties, e.g., molecular weight, diffusion coefficient, and charge density, resulting in erroneous decoding of symbols, especially when the Tx-Rx distance increases \cite{mcguiness2020analysis}. Accordingly, for error compensation, developing adaptive channel coding algorithms in conjunction with the modulation schemes is required.

\subsection{Detection}

Enhancing the detection performance of odor MC systems requires the development of advanced detection methods that effectively utilize the unique properties of odor communication across different channel and receiver combinations. In addition to advanced detection methods in the MC literature \cite{kuscu2022detection, 8445876, 8255665, li2019csi, 6708551}, detection schemes that consider cross-reactive sensing and combinatorial binding are also introduced \cite{jamali2023olfaction, civas2023frequencydomain}. By understanding the complex relationship between odor molecules and receptors, these approaches enhance the sensitivity and selectivity of odor detection systems, leading to improved overall performance. Several key design factors, including the detection threshold, receiver saturation, and sensitivity, significantly influence the development of these methods. By thoroughly analyzing and addressing these factors, we can overcome limitations and achieve improved detection performance in odor MC systems. Considering the time-varying behavior of the channel, estimating the channel state information is another challenge. Additionally, the number of distinct types of ORs, e.g., humans have $\sim\,$400 types of ORs, and the size of the olfactory alphabet, i.e., vectorial odor space, needs to be determined in an optimized fashion to generate odor mixtures for different communication scenarios, in which one or multiple TxRxs can be used. The remaining challenges, such as varying odor thresholds for detection and identification, the desired temporal relationship between odors and TxRxs, i.e., synchronization, the impact of continuous exposure on detection, i.e., Rx saturation, the presence of ambient odors, i.e., interference and background noise, and the effects of signal degradation, can all pose challenges to the detection mechanisms of odor MC systems.

Signal processing techniques can be used to enhance the ability of sensors to collect information in noisy environments, a quality that odor-based communications systems are often characterized by. Prediction models form a key aspect of operation since filtering alone fails when there is uncertain information about a dynamic system. Such odor-communication systems typically also possess Gaussian noise. For these reasons, the Kalman filter is exploited, e.g., it has been used to identify the location of a gas leak via a set of distributed sensors \cite{gasleakLiu}. In this manner, nature is inadvertently mimicked since it has been argued that the neural processing of sensory data can be understood as an implementation of Kalman filtering and is accepted as a mechanism from which consciousness could be considered to arise \cite{MillidgeKalman}. In addition to utilizing traditional signal processing methods, machine learning techniques have proven a valuable tool when employed in many applications to detect true outliers, e.g., the integration of Density-Based Spatial Clustering of Applications with Noise (DBSCAN)-based outlier detection and Random Forest in an IoT-based system to monitor several sensor signals in an automotive manufacturing factory \cite{10.1145/3593043}. 
In tandem with the advancement of electronic nose technologies, a growing body of literature exists in which machine learning techniques able to complement these technologies have been explored, e.g., for automated recognition of ‘fingerprints’ of volatile organic compounds that identify different phenomena to invoke selectivity and improve sensitivity \cite{Lujn2016MachineLM}.

\subsection{Open Issues}

In addition to the previously mentioned key design factors, such as detection threshold, receiver saturation, sensitivity, channel state information, and odor receptor diversity, there are several other future directions that can further enhance the performance of OMC techniques:
\begin{itemize}
\item \textit{Multi-Sensor Fusion}: Integrating various types of sensors, including chemical sensors, electronic noses, and biosensors, can improve the ability to detect odors by increasing sensitivity and selectivity. By combining data from different sensors to overcome challenges in sensitivity and detection precision, it becomes possible to accurately determine the threshold for detecting odors and effectively differentiate between different odors or odor mixtures.
\item \textit{Integrating Technologies}: By incorporating machine learning techniques into adaptive detection algorithms that combine traditional methods, it is possible to effectively tackle the challenges related to channel state information and the diversity of ORs. Machine learning leverages its ability to account for the dynamic nature of the channel and the wide variety of odorant receptors, enabling accurate estimation of channel state information, optimization of detection parameters, and substantial improvement in the overall performance of odor MC systems \cite{ye2021recent}. Furthermore, incorporating odor sensing systems with other advanced technologies like artificial intelligence, the Internet of Things, and data analytics amplifies their capabilities and range of uses. Developing smart sensor networks, cloud-based odor databases, and real-time monitoring systems that combine multiple technologies enables comprehensive odor detection and analysis.
\item \textit{Human-Machine Interaction}: It is particularly important for applications related to perception, cognition, and fragrance classification. It overcomes receiver saturation and sensitivity challenges by integrating human feedback and perception into odor detection systems, enhancing odor identification accuracy and reliability.
\item \textit{Odor Signature Analysis and Odor-Receptor Relationship}: Investigating the unique odor signatures, i.e., odor fingerprints, associated with specific odors, as well as understanding the relationship between odor perception and the physicochemical properties of odor molecules, can greatly enhance odor detection. Exploring the correlation between odor perception and molecular properties such as molecular weight, volatility, polarity, and functional groups will contribute to the development of more effective detection strategies \cite{hoehn2018status, genva2019possible, young2019smelling}.
\item \textit{Channel Access}: Another challenge is the efficient coordination of transmission strategies. Staggered transmission schedules, unique molecular identifiers, feedback-driven strategies, dynamic channel allocation, collision detection and recovery mechanisms, advanced algorithms for decoding, sharing, and coexistence techniques, and the introduction of biological feedback mechanisms all represent open issues. These issues encompass the need to enhance message discrimination through identifiers, optimize transmission based on feedback and changing conditions, allocate communication resources dynamically, manage conflicts and retransmissions, improve message decoding accuracy, ensure fair access in multi-transmitter scenarios, and establish bidirectional communication through biological entities, i.e., biological triggers \cite{khalid2019communication}. Addressing these complex challenges not only holds the potential to pave the path towards advanced communication systems inspired by biological processes but also to harness the unique attributes of odor mixtures for the creation of communication methods that are both highly efficient and adaptable.
\item  \textit{Standardization and Communication Protocols}: By establishing common communication standards, different odor detection devices and systems can seamlessly interact and share information, facilitating collaboration and enabling the development of integrated solutions in the field of odor detection.
\end{itemize}

\subsection{Quality of Smell for OMC}

Developing and defining metrics for evaluating the quality of smell and the quality of experience in OMC is crucial for advancing and standardizing OMC systems. These metrics serve as benchmarks to assess the performance, reliability, efficiency, and user satisfaction of OMC systems. By defining these metrics from both user and technical perspectives, researchers can address crucial questions, identify challenges, and focus on areas for improvement in odor communication technologies.

Reliability measures must be defined considering the subjective nature of the sense of smell, similar to metrics such as Signal-to-Noise Ratio (SNR), Bit Error Rate (BER), and Peak Signal-to-Noise Ratio (PSNR) used in digital communication and image processing. The goal is to establish a metric that captures the perceptual reliability of smell, considering the persistence and stability of odor signals in OMC. While persistent odors can be advantageous for certain applications, they can also introduce interference in the communication channel, posing challenges. For example, durability can be a more desirable metric in OMC compared to data rate requirements in digital communication. Determining the appropriate design criteria, including OMC transceivers, depends on the specific application and remains an open issue. Therefore, it is necessary to develop task-oriented communication evaluation metrics that focus on evaluating the received information in an intention-dependent manner rather than relying solely on bit-based information \cite{xu2023guest}.

Considering and defining the Quality of Experience (QoE) in OMC systems, parameters that affect user perception need to be taken into account. This includes adapting low-level abstraction parameters like delay and jitter when integrating OMC with digitized smell technologies and systems. High-level abstraction quality parameters such as richness, pleasantness, and intensity of the smell should also be considered to enhance the overall QoE \cite{alja2021qoe, gulliver2006defining}. Mean Opinion Score (MoS) is a valuable metric that can be used to evaluate and assess user quality perception in OMC systems. By incorporating MoS into the evaluation process, researchers can gain insights into how users perceive the quality of smell and tailor OMC systems to better meet their expectations.

A superior level of service in OMC can be achieved by establishing universal metrics for statistical analysis, similar to the way copulas are utilized in image processing. These universal metrics would enable a standardized evaluation of OMC systems, ensuring consistent and reliable performance across different platforms.

Furthermore, by drawing parallels to different image types, it is possible to define different types of signals for OMC. Binary odor signals correspond to the presence or absence of specific odorants, similar to binary images that represent black and white pixels. Discrete odor signals convey distinct categories or qualities of odors, akin to discrete images depicting separate objects or shapes. Continuous odor signals capture subtle variations in perception, resembling continuous images that demonstrate smooth transitions between colors and shades. These hypothetical mappings provide a framework for a comprehensive analysis of odor quality at different levels, facilitating advancements in the understanding and application of OMC systems. Through continuous advancements in both technical and user-oriented aspects, researchers can drive the field forward and facilitate the widespread adoption of OMC.

\section{Applications of Odor Communication}

\begin{comment}
\noindent This section overviews the possible application domains of odour-based communications. 
Odour molecules in MC have the potential to enable many applications in environmental, agricultural, industrial, and medical domains.
Smell integrated applications have been developed to detect rotten foods, environmental problems, medical diagnosis and to improve the interactivity and effectivity of existing technologies.
In [4] Dmitrinko et. al. implemented the olfactory notifications to the automotive industry to observe the effectiveness of the olfactions in the users attention, mood and comfort. In [5], Sanaeifar examples the detection of rotten foods.  InScent is an example of wearable smell integrated design for taking personal notifications in real life [7]. Meta Cookie+ is an example of a changed food perception with the usage of augmented reality [8]. There exist many other examples of smell based technologies such as VR implementations, museum and clothing stores demand and satisfaction themed applications, cooking games, movie and theatre increased reality examples, photo browsing applications, etc [9]. 
\end{comment}

In recent years, there has been significant progress in unraveling the fundamentals of the olfactory system and the properties of odor molecules. This progress has enabled the development of innovative systems and applications based on machine olfaction. These advancements have led to practical solutions in various domains that include food and beverage quality control, environmental monitoring and security, medical diagnostics, and interactive communication technologies. This section overviews the existing odor-based applications while highlighting the immense potential for new applications that OMC holds for the future.

\subsection{Healthcare and Medical Applications}

Olfaction, while playing a vital role in information transmission and communication among organisms, offers valuable insights for health and well-being. Leveraging olfaction directly can prove advantageous in human hazard perception and recognition. Remarkably, dysfunction within the olfactory bulb has been identified as an early indication of various diseases \cite{ruan2012olfactory, attems2014olfactory, dan2023loss}. It is conceivable to impede disease progression and mitigate associated impairments by intervening in the olfactory system. Thus, understanding and harnessing the potential of olfactory mechanisms can hold great promise for advancing diagnostic and therapeutic approaches for these diseases. %Any impairment in the sense of smell can diminish human well-being and indicate potential future serious illnesses. 
In many cases, olfactory impairments may be irreversible, as the existing treatment methods cannot fully restore these disorders. This is where OMC systems come to the forefront for olfactory substitution, enabling individuals to potentially regain their sense of smell. For other diseases where olfactory impairment serves as an early symptom, the olfactory region and other parts of the body may be affected. In such cases, besides alleviating the olfactory disorder, targeting the affected region related to the disease specifically can also be beneficial. Next, we will delve into a detailed explanation of the aforementioned scenarios.

\begin{figure*}[!t]
\centering
\includegraphics[width=0.8\linewidth]{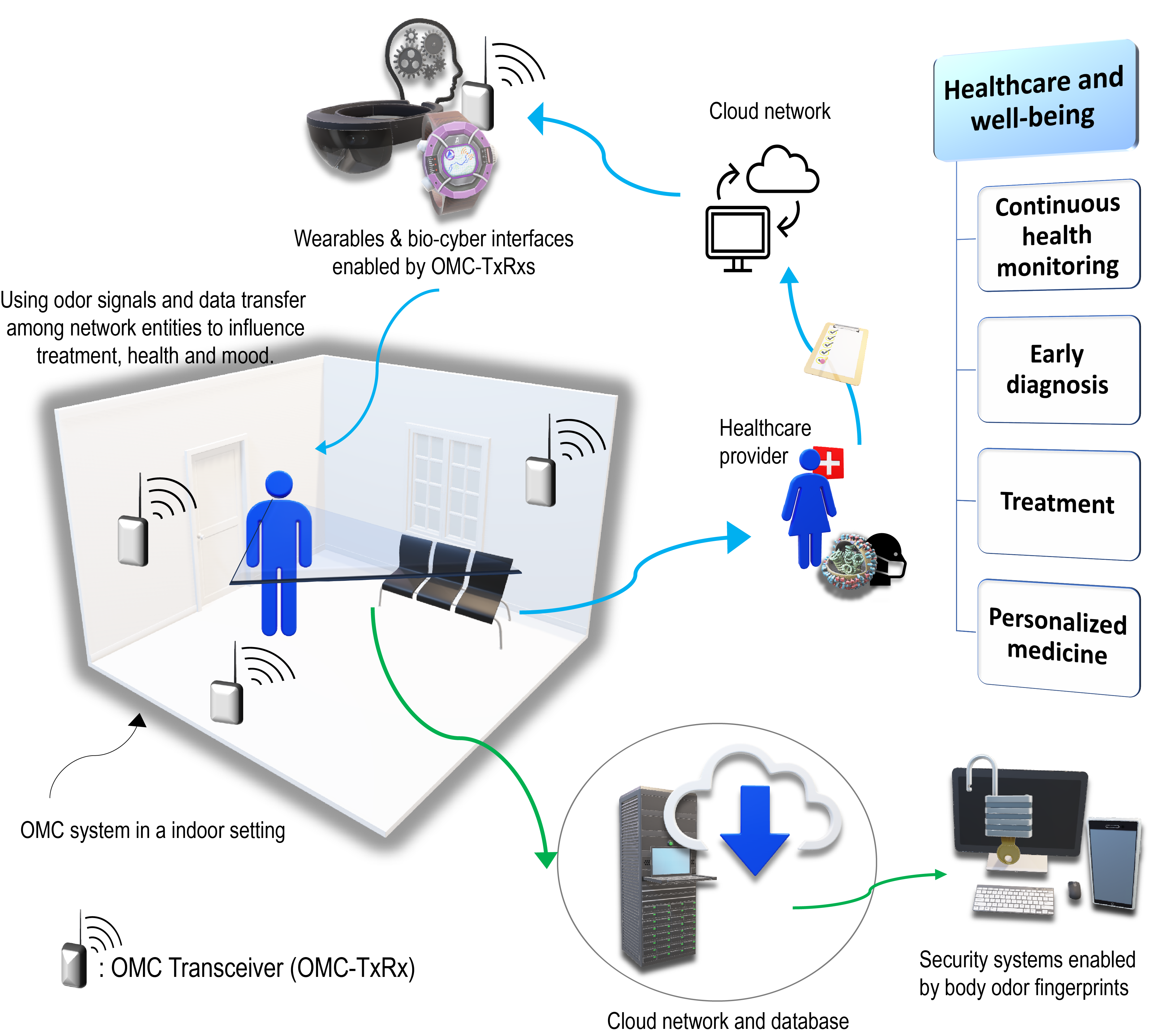}
\caption{OMC systems enable continuous analysis of breath and body odor in an indoor setting. Blue arrows represent the medical applications involving the collection and processing of volatile compounds from human breath. On the other hand, green arrows illustrate a specific security use case of OMC systems, which involves utilizing human body odor for biometric-based authentication \cite{naaz2022odortam}, creating databases, and obtaining odor fingerprints.
}
\label{App1}
\end{figure*}

\subsubsection{Health-care and well-being}
While the majority of our environmental perception is mediated through visual and auditory sensors, it is important to note that the olfactory system plays a smaller yet significant role in providing us with information about our surroundings. However, this does not diminish the importance of the olfactory system in preserving human health. The limited information we receive from the olfactory system can play a significant role in maintaining our health. The olfactory system serves as a warning to stay away from harmful substances and foods. It is evident that any damage to the sense of smell, such as anosmia, can lead to the loss of this critical ability to maintain health. Additionally, any disruption, such as hyposmia, severely reduces the olfactory system's ability to detect danger and alters it to an uncertain extent.  In situations where there is a loss or reduction in the sense of smell, using an alternative artificial method to detect odors can greatly contribute to preserving human well-being \cite{hurot2020bio, barbosa2018protein}. Designing and constructing systems for olfactory correction or olfactory substitution can be highly beneficial and vital for patients with olfactory disorders. Mechanisms based on next-generation OMC transceivers can effectively simulate the parts and performance of the olfactory system. Cases that involve sensitivity, such as faint toxic odors or odors that cannot be sensed by the human sense of smell, can be incorporated into the human olfactory system through the OMC mechanism. By employing this method, people can identify potentially dangerous sources with a satisfactory level of precision that might go unnoticed in normal circumstances. It appears that the implementation of OMC could not only partially alleviate the prevalent conditions of anosmia and hyposmia but also have the potential to significantly enhance the sense of smell. Enhancing olfactory capabilities can play a crucial role in preserving human well-being and preventing exposure to hazardous substances.
\subsubsection{Early diagnosis of diseases}
Besides its significance in preserving human well-being, multiple studies have indicated that the olfactory bulb is among the primary regions impacted by numerous illnesses. Presently, impairment of the olfactory bulb has been unmistakably proven during the initial phases of diverse neurological conditions \cite{thangaleela2022nasal, doty2012olfactory}. Interestingly, in certain neurological diseases, dysfunction of olfaction can occur even before the onset of symptoms. Thus, investigating the olfactory system shows potential for early detection of neurological diseases. Notably, Alzheimer's \cite{naudin2014olfactory}, dementia \cite{suh2020olfactory}, and Parkinson's disease exhibit this characteristic. In specific neurological disorders like Alzheimer's, dysfunction of the olfactory system can manifest prior to the appearance of typical symptoms. %The degeneration of the olfactory bulb may occur earlier or concurrently with the damage to the basal ganglia.  As a result, 
Early detection of olfactory disorders in their early stages using the continuous health monitoring system offered by OMC systems can be highly advantageous.
 
 %Therefore, it is strongly recommended that individuals undergo olfactory testing during routine annual check-ups so that any suspected disorders can be promptly referred to a specialist physician. 

In addition to neurological disorders, many infections, such as influenza, COVID-19, and similar cases, can also cause olfactory disturbances. However, these disturbances are temporary and not caused by damage to the olfactory bulb neurons. In these cases, differential diagnosis is important for distinguishing between infectious diseases and neurological disorders based on olfactory disturbances. In this regard, the accuracy of OMC systems can be significant in the diagnosis of these two groups of diseases. In the realm of neuronal diseases, the olfactory bulb neurons are susceptible to damage or loss, causing a disruption in the transmission and processing of olfactory information. Conversely, afflictions such as influenza may impede olfactory sensor function without impacting the intactness of transmitting and interpreting neurons. With the promising development of OMC techniques and devices for olfaction, there is a potential to stimulate olfactory sensors with greater efficacy. In such a case, if a subject experiences a comparable sensation to that of smell, they may have influenza, whereas the absence of olfactory interpretation could indicate a neurological disorder. Further differential diagnosis among neurological diseases can be pursued by early diagnosis systems enabled by OMC systems.
 
 %If it is possible to make a differential diagnosis, this method is completely non-invasive and does not require surgery. Also, like other methods, such as MRI, which uses a very powerful magnetic field, it will not have side effects. 
 
 %Another advantage of using MC tools in the field of diagnosis is that costs are reduced, and it can achieve the best results in a short time. 
 
The use of OMC cannot only be useful in early diagnosis but may also be effective in the treatment of diseases. After the diagnosis of a disease, attention can be turned to identifying the underlying causes of the disease. Some neurological diseases are caused by genetic disorders. Using OMC to eliminate genes in the early stages of the disease can prevent its progression \cite{blenke2016crispr}. As illustrated in Figure~\ref{App1}, OMC systems can enable continuous analysis of breath and body odor in an indoor setting \cite{khalid2019communication}, Through the detection of biomarkers, pathogens, and other particles in human breath, these systems facilitate various medical applications.  Furthermore, wearables, such as smart watches and glasses, and bio-cyber interfaces, empowered by OMC-TxRxs, can utilize odor signals and data transfer among network entities to influence mood, improve health, and provide treatment. In the following sections, we will delve into these medical applications, providing detailed explanations and insights.

 %\textcolor{purple}{For the various activities conducted in human health care, comprehensive information about the behavior of normal participants is needed. They exhibit various biometric behaviors that differ from each other. %Therefore, if we can separate these common characteristics, we can identify the set of healthy features and flag them% Therefore, by separating these common characteristics, the set of healthy features can be identified and labeled when even the slightest change from a healthy state occurs, enabling more precise medical examinations. To achieve this goal, it is better to obtain big data based on olfactory information. This is where all olfactory-related information is collected through a mobile application or any other device designed based on MC and gather it in a central server. In this way, the more information can collect about healthy individuals, the more accurate the identification of healthy individuals becomes.}

\subsubsection{Treatment}
Another cause of neuronal destruction can be the presence of certain toxins (such as MPTP for Parkinson's disease \cite{mustapha2021mptp}) in the environment. Certain toxicants, including nanoparticles, can damage the olfactory receptor cells and enter the brain through the olfactory mucosa, potentially affecting the structures of the central nervous system. This evidence of the adverse effects of toxicants on the olfactory system emphasizes the significance of maintaining a healthy olfactory function for everyday functioning and safety \cite{genter2019toxic}. The use of OMC systems can be useful in attacking these toxins and preventing their accumulation. Some detoxifying compounds can be directed toward the site of toxin accumulation. If combined with the targeted toxins, the nature of the harmful substances can be altered and transformed into harmless compounds for the body. Ultimately, these compounds are collected by the circulatory system and excreted by the renal system. Similarly, the accumulation of free radicals can also cause neuronal death. Therefore, it is possible to deliver free radical scavengers to them and transform them into harmless compounds through OMC systems. However, in the event of neuronal death in a specific region of the brain, an MC-based system can be developed to create a substitute for neuronal function so that some of the functions of damaged neurons can be performed \cite{dismukes1979new}. If any of these therapeutic approaches can be implemented, a significant step can be taken toward treating neurological diseases.

Current efforts in molecular communication for healthcare \cite{koca2021molecular, felicetti2016applications, chahibi2017molecular, CHAHIBI201790, atakan2012body, khan2020nanosensor, malak2014communication, barr, abnor, cevallos2019health} serve as a foundation for the development of OMC systems, where odor molecules carry specialized information to diagnose and treat olfaction-related diseases. Early detection and monitoring systems for olfactory disorders enable timely intervention and personalized treatment options. By implementing these OMC applications, healthcare providers can significantly improve the outcomes of individuals with olfactory impairments, as seen in Figure~\ref{App1}. These technologies can play a vital role in enhancing their quality of life by ensuring prompt interventions based on individual needs.

\subsubsection{Olfactory memory}

Knowledge regarding odor has the potential to greatly impact how our brain processes different stimuli. The cognitive functioning of individuals can be influenced by their sense of smell, as a designated area in the brain stores memories related to odors \cite{gao2021artificial}. %The information obtained from olfaction and olfactory memory can enhance our ability to detect and recognize stimuli in our environment. It is evident that individuals with olfactory disorders experience cognitive impairments that affect them. The olfactory system, which encompasses the olfactory epithelium, olfactory receptors, olfactory bulb, and olfactory cortex, is essential for detecting and processing odors. The olfactory epithelium's receptors detect odor molecules and transmit signals to the olfactory bulb. 
OMC systems enabled through bio-cyber interfaces \cite{akyildiz2019microbiome, aktas2022weight, cyber1, el2020mixing, spinal, sinw, iobnt} have the potential to play a significant role in harnessing olfactory memory for healthcare and medical applications such as odor-based therapy or interventions, as illustrated in Figure~\ref{App1}. Olfactory memory can have implications for diagnosing certain medical conditions. Olfactory memory assessments can contribute to the identification and evaluation of medical conditions, including Alzheimer's disease \cite{Croy2015Test-retest,Kollndorfer2017Assessment}.

%It is now feasible to record and analyze odor signals in a digital format by fusing cyber elements, such as sensors and data processing algorithms, with biological olfactory systems. 

%The study of applications in scent-based communication is made possible by this interface's facilitation of the storing, retrieval, and transmission of olfactory information. 
Accordingly, these bio-cyber interfaces can enable the study and exploration of applications in OMC by facilitating the storage, retrieval, and transmission of olfactory information.
Understanding the molecular mechanisms underlying olfactory processing and memory formation can facilitate the development of interventions and treatments for olfactory disorders. %The utilization of MC can be beneficial in both olfactory recognition and olfactory memory. 
Our ability to perceive and interpret external stimuli relies on biological receptors, which possess the remarkable capability to detect a vast array of sensory information and convert it into signals that can be processed by our brain. Similar to molecular communication, which relies on the use of molecules for transferring information, these biological receptors serve to enable the transmission of information from our surroundings into our brain. This allows us to comprehend and interpret the world that surrounds us. Understanding the molecular mechanisms underlying olfactory processing and memory formation can aid in the development of interventions and therapies for individuals with olfactory disorders. 

\subsection{Human-to-human Communication}
The diversity of odor communication that supports the functioning of natural ecosystems extends to unconscious human-to-human signals. While there are also clearly beneficial ways in which these natural phenomena can be exploited, the ability to manipulate cognition or emotional states must be understood, and protocols must be set up to ensure this advancing field of research is explored in an ethical way. These human-to-human interactions can take different forms:
\begin{itemize}
\item\textit{{Facilitating bonding:}} Volatile compounds are known to play an important role in mother-infant bonding and, hence, brain development at a point where the visual system cannot act as a mode of identification \cite{svaglio2009mother}. Additionally, a compound in tears has been found to reduce male production of testosterone and sexual arousal. Tears are known to possess a different composition depending on the emotional state of the individual and consequently invoke a different response \cite{Gelstein2018}.
\item \textit{{Invoking physiological changes that better prepare the individual to their environment:}} It has been proposed that the phenomenon in which a pheromone invokes menstrual synchrony in females living in proximity could be made use of in the design of inhalable contraceptives \cite{svaglio2009mother} As a second example, testosterone exposure to pregnant women is argued to influence fetal development in a plethora of ways, including influencing the development of autism, ADHD, sexuality, and ranging to the ratio of finger lengths \cite{bailey2005finger, balthazart2018}. Pheromones, which pose more challenges to monitor, are also likely to influence fetal development in ways that could be evolutionarily favoured.
\item \textit{{Eliciting sexual disgust or appeal:}} The degree to which the immune system of a person is similar influences how pleasant their bodily aroma appears. It is a phenomenon that arises due to the genetic advantages this translates to the offspring and is known to influence partner choice. This response is known to not occur in women who are on hormonal contraceptives \cite{Sorokowska2018}, indicating the power and therefore caution that must be exercised in chemical intervention for unconscious human-to-human communication processes.
\end{itemize}
The influence of odor on unconscious human interactions can be further investigated through the adaptation of OMC systems. OMC transceivers, which can take the form of mobile devices, outdoor sources, implants, or wearables, have the ability to trigger unconscious states in individuals in both indoor and outdoor settings. This understanding can have implications in various fields, including contraceptive design, where incorporating pheromonal effects and odors into inhalable contraceptives may potentially influence partner choice. The potential benefits of these applications need to be realized while also respecting individual autonomy and upholding societal values.

%% Please feel free to make changes here! I'm looking at how I can make the style a bit more cohesive with yours. - Melanie 

\subsection {Industrial and Environmental Applications} 
The sense of smell, often overlooked compared to other sensory perceptions, plays a significant role in various industries. Odor applications in industry encompass a wide range of sectors, including environmental monitoring, food and beverage production, marketing, autonomous navigation, and monitoring of toxic agents and pollutants, which are reviewed in this section. Understanding and harnessing the power of human odor detection technologies can lead to advancements in these areas, improving processes, quality control, and overall customer experience \cite{li2014recent}.

\subsubsection{Food Industry}

Odor sensing technologies have attracted enormous attention in the food industry since volatile organic compounds are important indicators for the quality and safety of food products. Odor imaging technology is the most promising method for analyzing food products. The accurate evaluation of food odors and ensuring the quality and safety of food products are crucial factors as they have a direct impact on human health. As mentioned before, odor molecules are widely used in healthcare and influence human perception. Hence, controlling food odor enables the creation of a different perception of food consumption that can be further used in various diets to help patients avoid consuming the foods that they are addicted to or suffer from. Conversely, the odor of food is intricately associated with personal taste and consumption patterns. Consequently, the fragrance of food exerts a significant impact on both the demand for and sales performance within the culinary sector.

\subsubsection{Environmental Monitoring}

The identification, measurement, and reduction of odor emissions can all be facilitated through the use of OMC systems in environmental monitoring. Industries may efficiently manage and handle odor-related concerns, delivering a healthier and more sustainable environment for all stakeholders by using human odor-detecting systems and engaging the community in odor reporting. Odor emissions from industrial facilities, landfills, or agricultural operations may substantially negatively influence the environment and the populations in the area. Effective monitoring and control of these scents are crucial for maintaining public health, minimizing complaints, and promising regulatory compliance.

%Trained odor assessors and E-noses are only two examples of human odor-detecting technology that evaluate and quantify odor emissions to help with odor source identification and effect evaluations. E-noses use sensor arrays and pattern recognition algorithms to detect and identify particular odor compounds, enabling continuous monitoring and real-time data collection. Trained odor assessors use their sensitive olfactory senses to assess odor intensity, character, and offensiveness.

%The establishment of odor panels or community odor boards is another result of improvements in odor communications. Community members are actively encouraged to report and record odor occurrences using these channels. By interacting with the local community, industries may learn important information about the perceived odor issues, the frequency and length of odor occurrences, and the possible health effects on inhabitants. Transparency, accountability, and improved communication between industry stakeholders and the community are fostered by this collaborative approach.

\subsubsection{Safety and Security}

The detection of specific odors linked to illegal substances holds considerable importance in areas such as law enforcement, transportation security, and public safety. Traditionally, individuals with specialized training, like sniffer dogs or human experts, have been relied upon to identify the smells of explosives and drugs. Nonetheless, there have also been advances in the development of electronic devices such as E-Noses that serve this purpose. Integrating future OMC systems with these tools provides an additional layer of protection by enabling the detection and identification of even minute quantities of explosives or illicit narcotics. This combination enhances the effectiveness of security measures by assisting in preventing potential threats and ensuring public safety. As an example, one application in the realm of smart cities for enhancing safety and security is the incorporation of odor sensing technologies, which can detect and identify potentially dangerous or suspicious odors. This adds an extra layer of monitoring and threat prevention \cite{troisi2022managing}.

%Furthermore, E-nose devices, mentioned earlier, are also being explored for safety and security purposes. Thanks to their sensor arrays and sophisticated pattern recognition algorithms, these devices can identify and categorize a wide range of scents, including those linked to dangerous compounds. They provide benefits including rapid reaction times, non-intrusive sampling, and the capacity to monitor vast regions successfully.

\subsubsection{Product Development and Marketing}

Technologies for odor detection are increasingly being used in product development and marketing to better understand customer preferences, enhance sensory experiences, and produce enticing perfumes. Consumer goods, including fragrances, cosmetics, and personal care items, are one area in which odor detection technology is used to create new products. Companies may assess and examine the olfactory profiles of various formulations by using trained sensory panels or E-Nose devices \cite{bios11030093}. Technologies for odor detection are used in the food and beverage sector to improve flavor profiles, product quality, and consistency. To achieve the desired sensory characteristics, professionals with expertise in evaluating odors and off-tastes play a crucial role in guiding the creative process. In addition to E-Noses, market research techniques such as sensory panels and consumer surveys that involve human assessors are still commonly employed. These methods allow trained experts with heightened olfactory senses to analyze odor profiles and offer their evaluations of different perfumes. The ability to comprehend customer preferences and make informed decisions regarding fragrance selection hinges on these assessments.

\subsubsection{Gaming, Education and Automotive}

Odor technology plays a vital role in various industries, including the gaming, education, and automotive sectors, considering its ability to influence attention and memory. In gaming, integrating scents into games enhances player engagement, prolongs game sessions, and creates a more immersive experience. The direct impact of odor molecules on human moods and memory triggers makes scent integration an effective tool for developers to captivate gamers \cite{olofsson2017beyond}.

In the automotive sector, there are exciting opportunities to improve safety and driving experiences through odor applications. By incorporating attention-enhancing odor-release systems enabled by OMC systems into vehicle ventilation and air conditioning, accidents can be prevented by enhancing driver alertness and focus. Additionally, odor stimuli can provide drivers with task-related information, effectively utilizing odors as an additional sensory channel for conveying navigation instructions or alerts about road hazards.

By integrating odor technology into educational and training settings, there is a significant potential to improve focus and retention. This can be achieved by creating learning experiences that engage multiple senses, trigger emotional responses, and establish connections with the surrounding environment. Through harnessing the capabilities of odors, educators have the ability to enhance cognitive abilities in students by fostering an innovative approach that incorporates contextual associations along with multi-sensory learning techniques.

%Human perception has a place in developing various applications in different industries, such as video games and automotive industries. For example, there are vast efforts to generate smell-integrated video games and applications since odor molecules directly affect people's moods and can evoke memories that can further increase the willingness of people to continue playing the game or increase the sense of pleasure while playing. This ability of odor molecules to influence the mood and behavior of humans also promotes different applications within the automotive industry. For instance, traffic accidents can be prevented with attention-enhancing odor release systems that can be placed in the ventilation and air conditioning systems of vehicles. In addition, task-related information could be provided to drivers using olfactory stimuli.

\subsubsection{Agricultural technologies}
Volatile compounds are the channel through which plants may communicate most quickly, and therefore they are the exclusive mode of communication where time-dependent concerns must be conveyed. 
Each year, between 20-40\%  of crops are expected to be lost worldwide to disease \cite{nifa2023}. Rice and other grains are particularly susceptible to the recurring blast epidemics, which claim 10-30\%  alone. Jasmonic acid, a plant stress hormone, is released by a wide variety of plants in response to damaging conditions, including mechanical damage from pests, UV damage, and insufficient water, among other stressors. The specifics of the damage are often conveyed via different compounds that can be directly linked to a metabolic process. Consequently, an all-purpose and damage cause identifying OMC systems may be developed to monitor the health of crops so that preventative measures can be taken to isolate the damage \cite{enose2022polyimide}. 
Vertical farming can also be a prime subject for the application of OMCs due to the confined nature of this practice. The advent of this mode of farming enables the collection of higher concentrations and superior accuracy of the information collected on volatile compounds while also minimizing noise. This permits ML algorithms to be recruited to optimize the conditions required for the plants or yield, which may be achieved by reading signals and responding to the plants needs or via signal mimicry \cite{enose2022polyimide}.
Optimum harvest time may be recognized via either the volatile compounds emanating from the fruit or plant signaling. The release of spores and ‘pollen coupling’ \cite{satake2000pollen} are also coordinated in ways that can be exploited to increase pollination frequency and maximize yield.
Genetic engineering offers a tool to modify the biological receiver's behavior in response to chemical triggers. In agricultural technologies, this could find use in attracting or repelling insects or involving methods to increase fruiting frequency. Following modification, plants could themselves act as repeaters or as a method of mapping the location of the pest attack.

\subsubsection{Defence systems}
Research in scent-based communication for defense systems offers significant benefits in three main areas: signal encoding, threat location, and weapon identification. As a new mode of signal encoding, and depending on the abilities of the receiver to distinguish molecules, signal interception may prove difficult given the number of available vectors that chemical structure can encompass. In the area of threat location, by understanding the mechanical damage hormones from different plants and creating a 2D map of the vegetation, disturbance along potential pathways may be recognised via herbivore-induced plant volatiles (HIPVs). Another area that can benefit from OMC-enabled odor applications is weapon identification. While chemical weapons may exist in very low quantities, the resulting disruption of plant or organism metabolic pathways might be more significant and easier to recognise. In this manner, we may indirectly measure the consequences of an attack or testing. This concept is supported by prior research in which genetically engineered plant-microbe interactions have allowed contaminants to be identified \cite{toth2022microbe}.

\begin{figure*}[!t]
\centering
\includegraphics[width=0.9\linewidth]{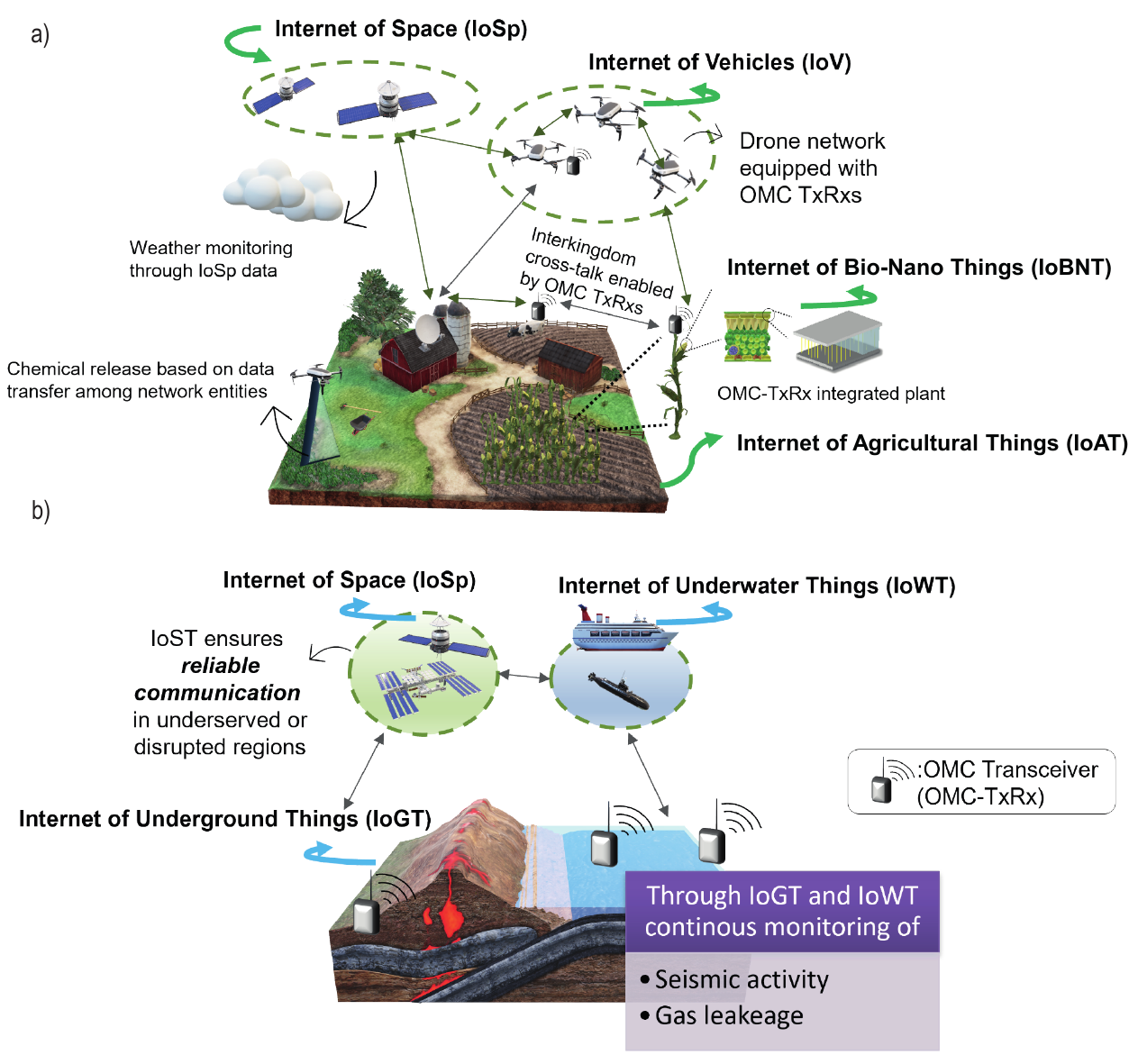}
\caption{Empowering the Internet of Everything (IoE) concept through OMC entities a) OMC-TxRs integrated into the Internet of Agricultural Things (IoAT) for smart agriculture and precision farming, connecting with Internet of Space (IoSp), Internet of Vehicles (IoV), and the Internet of Bio-Nano Things (IoBNT). b) Harnessing Internet of Space (IoSp), Internet of Underwater Things (IoWT), and Internet of Underground Things (IoGT) connections for climate and earthquake monitoring systems.}
\label{App2}
\end{figure*}

 \subsection{Broader Vision on Odor Applications}

Odor communication possesses immense potential due to its persistent broadcast nature. The long-lasting odor signals can be considered as letters left in a mailbox to be collected when convenient. This is a potent 
capability compared to that of other communication modalities, both conventional and unconventional, as odor messages can linger in the transmission medium, enabling extended message lifetimes. This feature opens up intriguing possibilities across multiple domains, such as \textit{agriculture}, where persistent odor molecules can offer useful insights into crop health, soil conditions, or pest presence. They can also help study animal behaviors, migration patterns, and habitat changes. Persistent odor signals are also promising for \textit{spatiotemporal tagging and marking}, which is crucial for especially search and rescue operations. For such efforts, odor signals can serve as enduring cues, assisting navigation even after the first signal transmission and helping individuals or teams locate people reported as lost or missing. Another tagging and marking opportunity is \textit{odor-based landmarks}, which can provide reliable reference points or directional cues in intricate environments, both outdoors and indoors, revolutionizing navigation. Odor messages overkept on the channel can also help preserve historical artefacts and cultural heritage. By encapsulating scents associated with specific events, locations, or eras, \textit{olfactory archives} can be formed, offering a multi-sensory and immersive experience for future generations to connect with the past. The multi-sensory interaction concept can also be introduced into \textit{virtual/augmented reality} (VR/AR) applications and \textit{online gaming} to offer enhanced immersion to users by adding another sensory dimension to the existing settings. In this way, scents can be experienced corresponding to the virtual environment, enabling users and players to interact with each other through odor-based cues. Despite the huge potential, this unique capability is subject to various external factors, such as air circulation, diffusion, and competing or dominant odors in the transmission medium, which can adversely impact the detectability and longevity of odor molecules. Hence, medium-resilient odor formation, coding, and modulation techniques are of utmost importance.

\subsubsection{Digital Scent Technology and Multimedia}

The OMC system, introduced in this paper, presents a novel approach to particle-based communications by utilizing odor molecules as carriers of information. This distinguishes it from digital scent technologies that focus solely on digitizing smells and rely on controlled computer devices for odor-related transmissions. Therefore, digital scent technologies can be considered sub-applications within the broader OMC system.

In near future, it is conceivable that movies and video games will go beyond the mere transmission of visuals and sounds. This transformation would involve the integration of additional senses like taste, achieved through the digitization of chemical senses \cite{spence2017digitizing}. Pioneering efforts are underway to develop electrical tongue devices and other innovative technologies aimed at simulating taste sensations \cite{ranasinghe2012digital, ranasinghe2016digital}. These devices employ various mechanisms, such as thermal changes and electrical stimulation, to recreate the intricate flavors. The integration of these devices into everyday electronics can enhance the immersive multimedia experience, enabling users to not only see and hear but also taste and feel digital content. This progress represents a significant leap toward a future where our senses are fully engaged and seamlessly integrated into the digital world, offering endless possibilities for entertainment, communication, and beyond.

Furthermore, this integration of additional senses goes beyond taste alone. It encompasses the dissemination of corresponding smells and tactile feedback, fostering a more comprehensive and immersive multimedia experience. In this hypothetical scenario, electronic devices, including televisions, smartphones, and computers, would be equipped with embedded systems capable of generating and transmitting both smells and tactile sensations. To achieve this vision, advancements in the field of the tactile internet are paramount. The tactile internet refers to a network infrastructure that enables the seamless transmission of tactile sensations, allowing users to perceive and interact with remote objects in real-time \cite{Mumtaz2019Guest, fettweis2014tactile}. By integrating the methods and technologies of the OMC system with the tactile internet, the transmission of tactile sensations can be enhanced alongside odor information.

Computer-controlled olfactory devices, capturing and transmitting odor information, allowing for the faithful recreation of specific scents at the receiving end, aimed in digital scent technology \cite{ghinea2011olfaction}.  OMC transceivers offer the potential to make these devices a reality, even though they are not yet available commercially. The integration of existing digitized smell technologies seamlessly with OMC transceivers underscores the importance of universal transceivers \cite{civas2021universal}. By establishing standardized protocols, the OMC system facilitates interoperability between different communication systems and devices, ensuring successful transmission and reception of smells and tactile feedback across various platforms. Not only does the OMC system revolutionize olfactory communication by using odor molecules as carriers of information, but it also opens up possibilities for integrating additional sensory modalities into digital communication systems, potentially in a hybrid fashion.

\subsubsection{IoE enabled OMC systems}

The Internet of Things (IoT) is a collection of various connected devices that are able to exchange data with each other and with applications through the internet. This includes smart cities, industrial systems, agricultural networks, and wireless sensor networks. However, there is often a lack of interaction between these different branches or sectors within the IoT. To address this issue, the concept of the Internet of Everything has emerged \cite{beyond2019internet}. IoE aims to establish new pathways for interaction among existing sectors within IoT \cite{IoE}. This holistic approach enables more seamless integration and communication between different parts of the IoT. In the context of OMCs, where odors are used as information carriers for data transmission, incorporating odor-based communication technologies into the IoE vision holds great potential for unprecedented applications. This can involve data transfer between different IoXs using not only particle-based communication modalities but also engaging other modalities such as electromagnetic waves, optics, and acoustics.

One fascinating concept within the IoE vision is the idea of connecting interkingdoms, which involves establishing communication and interactions among different realms of life. This concept envisions integrating olfactory communication within plants and between plants and other entities, such as animals and human-made objects, as illustrated in Fig. \ref{App2} a). By leveraging odor signals, plants can communicate their conditions and interact with their environment in a more sophisticated manner.

Through the interaction pathways of the Internet of Agricultural Things (IoAT), Internet of Vehicles (IoV), and the Internet of Sensing Things (IoST), smart farming and precision agriculture can be enabled. In this context, OMC transceivers embedded in drones or mobile devices can gather agricultural information in an OMC network and transmit the collected data to cloud platforms. Furthermore, the manipulation of specific odor molecules or mixtures to trigger desired responses in plants, such as promoting growth, enhancing nutrient uptake, or regulating reproductive processes, can be achieved through envisioned OMC systems connected with IoXs, e.g., via drones equipped with OMC-TxRxs, as seen in Fig. \ref{App2} a). This includes the integration of the Internet of Bio-Nano Things (IoBNT), where nanosensor networks are integrated into the biological systems of plants \cite{civas2021universal, mou2022efficient, dixon2021sensing} to control monitoring and data transmission from nano-scale biological entities to cyber domains. By combining this data with regularly updated geospatial information and local weather predictions, facilitated through edge computing in the context of the Internet of Space (IoSp), smart farms can take proactive measures to mitigate risks and optimize their operations for sustainable practices \cite{dixon2021sensing}.

Climate and earthquake monitoring systems can be another use case of IoE concept enabled by OMC. Fig. \ref{App2} b) illustrates the potential application of IoE in continuous monitoring of seismic activity and gas leakage within the Internet of Underwater Things (IoWT) and Internet of Underground Things (IoGT) systems \cite{saeed2020climate}. This monitoring capability significantly contributes to sustainability efforts and enhances disaster management. Additionally, the interaction of Internet of Space (IoSp) entities enables reliable communication in underserved or disrupted regions, further strengthening the overall effectiveness of the system.

Several key technologies, such as digital twin, metaverse, augmented reality (AR), semantic communications, compressive sensing, 6G, and the Internet of Senses (IoS), serve as enablers of the IoE. Through their integration, the sense of smell can be expanded through odor-based bio-cyber interfaces enabled by OMC systems. These technologies play a crucial role in facilitating seamless integration and communication between different branches or sectors within the IoE. This integration has the potential to create various interactions and applications. One example is within the Internet of Bio-Nano Things, where OMC transceivers in an OMC nanonetwork can be used to stimulate the sense of smell, leading to improvements in mood, learning, and olfactory memory retention. Moreover, when it is combined with the Internet of Battlefield Things (IoBT), monitoring and influencing soldiers' physiological and emotional conditions can be enabled. In relation to the Internet of Vehicles, integrating OMC systems can effectively capture driver attention through odor stimuli, which ultimately enhances road safety and awareness. Moreover, cloud domains that process odor fingerprint databases collected via IoST networks can open up possibilities for security applications focused on identity verification and authentication based on individuals' unique odors \cite{naaz2022odortam}.

\noindent 
 %Privacy için kriptoloji temelli bir kokunun korunarak iletilmesi yöntemi önerilebilir.

\begin{comment}
E-Nose Applications
Dung et. al reviews the bio-noses scope of applications such as lung cancer detection, clinical diagnosis, ingredient and fermentation control of foods, cosmetics, brand recognition and malodour monitoring [6]. Main drawbacks of the e-nose applications are the purification of odour molecules, the difficulty of getting it dispersed through the air, the need to use animals in the experiments and the difficulty of the odour categorization [6].

\end{comment}

%\input{Figures/Figure4.tex}

\begin{comment}
    The development of multisensory technologies, such as virtual reality combined with digital scent technology (olfaction), can merge the Internet of People and Senses with the Metaverse, creating unprecedented opportunities such as the ability to "teleport" to digital twins and enhancing the sense of reality in the virtual environment. For instance, Panagiotakopoulos et al. demonstrated that people stimulated by the electrical signals in the nose (to stimulate olfaction) could find the odor source in the virtual environment better than the real-life [58,59]. However, digitizing and recreating some senses are more complex. For example, digitizing and synthesizing scent is still an open issue, while digitizing and recreating sound and sight can be achieved. 
\end{comment}

\balance

\section{Conclusions}

In conclusion, olfaction is a fascinating sensory system that holds great potential for communication engineering and various applications in the IoE domain. Understanding the principles of olfactory signaling can enable the development of artificial and controlled odor-based information transfer systems between humans and human-made things, improving human performance and mental and physical states. Despite their huge potential, the principles of odor-based information transfer have not yet been thoroughly investigated from an ICT perspective. Thus, this paper aimed to review the state-of-the-art, challenges, concerns, and future research directions in the field of odor communications, including designing odor MC transceivers, developing odor MC techniques, and modeling an end-to-end OMC channel. Finally, we highlighted the potential application areas of communication engineering in odor-based communications, including early diagnosis of various diseases and a better understanding of chemical perception. Overall, this paper can guide researchers in achieving the ambitious goal of taming and engineering olfactory communication as a new communication modality in IoE.

%%if part of the goal of the paper is to win funding, then as a final word we could emphasise the importance of the objectives we are planning to list there (and beneficiaries). E.g. "We have identified areas in which the adoption of principles from odor based communications may prove transformative... The keys steps in realising these are first the development of e-nose technologies capable of better recognising larger molecules and the molecular shape. This will transform the use and understanding of odorants which leads to the second key step which is to..."

\bibliography{References}
\bibliographystyle{IEEEtran}

\end{document}